\documentclass[pra,
 reprint,
 amsmath,amssymb,
 aps,
 superscriptaddress,
 floatfix,
 nofootinbib,
 natbib,
 longbibliography
]{revtex4-2}

\usepackage{graphicx}
\usepackage{xcolor}
\usepackage{amssymb,amsmath,mathtools,mathrsfs,slashed}
\usepackage{amsfonts}
\usepackage{dsfont}
\usepackage{braket}       
\usepackage{dcolumn}
\usepackage{bm}
\usepackage{hyperref}
\usepackage[normalem]{ulem}

\usepackage{algorithm}
\usepackage{algpseudocode}
\floatname{algorithm}{Protocol}

\algrenewcommand{\algorithmiccomment}[1]{\hspace{-1.2cm} #1}

\definecolor{mymagenta}{RGB}{200, 0, 100}
\definecolor{myblue}{RGB}{45, 48, 146}
\definecolor{mygreen}{RGB}{0, 126, 0}
\definecolor{myorange}{RGB}{255, 136, 19}
\graphicspath{{figures/}}

\begin{document}

\preprint{APS/123-QED}

\title{Shallow-depth GHZ state generation on NISQ devices}

\author{S.~Siddardha Chelluri}\thanks{S.S.C. and S.S. contributed equally to this work. Correspondence should be addressed to S.S.C. (schellur@uni-mainz.de) and S.S. (stephan.schuster@fau.de).}
\affiliation{Institute of Physics, Johannes-Gutenberg University of Mainz, Staudingerweg 7, 55128 Mainz, Germany} 
\author{Stephan Schuster}\thanks{S.S.C. and S.S. contributed equally to this work. Correspondence should be addressed to S.S.C. (schellur@uni-mainz.de) and S.S. (stephan.schuster@fau.de).}
\author{Sumeet}
\affiliation{Department of Physics, Friedrich-Alexander-Universit\"at Erlangen-N\"urnberg (FAU), 91058 Erlangen, Germany}
\author{Riccardo Roma}
\affiliation{Theoretical Physics, Saarland University, 66123 Saarbr\"ucken, Germany}
\affiliation{Institute for Quantum Computing Analytics (PGI-12), Forschungszentrum J\"ulich, 52425 J\"ulich, Germany}

\date{\today}

\begin{abstract}

In this work, we focus on GHZ state generation under the practical constraint of limited qubit connectivity, a hallmark of current NISQ hardware. We study the GHZ state preparation across different connectivity graphs inspired by IBM and Google chip architectures, as well as random graphs that reflect distributed quantum systems. Our approach is a measurement-based protocol designed to utilize qubit connectivity constraints for the generation of GHZ states on NISQ devices. We benchmark this against a tailored version of state-of-the-art unitary-based protocols, also incorporating physical connectivity limitations. To evaluate the performance of the protocols under realistic conditions, we conducted implementations on the IBM Eagle r3 chip. Additionally, to explore near-term scalability, we performed simulations across a range of graph sizes and connectivity configurations, assessing performance based on circuit depth, the number of two-qubit gates, and measurement overhead. We observe a trade-off between the two protocols across different figures of merit. For current state-of-the-art NISQ architectures, the unitary-based protocol is more suitable, as it avoids mid-circuit measurements and classical feedforward. However, the measurement-based protocol is expected to become more advantageous in the future with more error-resilient quantum devices, owing to its reduced circuit depth and consequently shorter execution times. In both settings, our proposed method provides an efficient means of leveraging the topology of qubit connections available on a given device.

\end{abstract}

\maketitle

\section{Introduction}

Efficient generation and control of multipartite entanglement is a fundamental requirement for implementing a wide range of quantum information protocols. Among multipartite entangled states, Greenberger-Horne-Zeilinger (GHZ) states are particularly notable for exhibiting genuinely nonlocal correlations~\cite{Pan2000GHZ}. These states are an essential resource for quantum internet ~\cite{Kimble2008, q_net}, including quantum key distribution (QKD)~\cite{BB84,e91}, as well as for quantum sensing~\cite{Hyllus_2012, Pezze2014, Kielinski_2024} and distributed quantum computing \cite{CCT+20,dqc_survey, dqc_ion}. With the rapid development of quantum hardware across several platforms, the ability to efficiently generate high-fidelity GHZ states at scale is a key milestone for near-term, noisy intermediate-scale quantum (NISQ) devices. However, in the NISQ era, generating GHZ states remains a fundamental challenge, as resource requirements often scale poorly with system size, leading to deeper circuits, more entangling operations, and higher sensitivity to noise, gate errors, and decoherence \cite{Preskill2018}. 

Numerous methods have been proposed for generating GHZ states on digital quantum computing hardware~\cite{Cruz_2019, Baumer2024, Liao2025, Kuan-Cheng, Moses2023, de_Jong_2024, Bao_2024}. Linear depth constructions are straightforward but result in deep circuits that are especially vulnerable to noise in systems with limited qubit connectivity~\cite{Cruz_2019, Baumer2024, Liao2025, Kuan-Cheng}. In contrast, logarithmic-depth circuits and measurement-based protocols can mitigate circuit depth, but they typically assume idealized qubit connectivity resulting in additional operations that are costly to implement on current quantum hardware~\cite{Cruz_2019, Baumer2024, Moses2023}. Bäumer et al.~\cite{Baumer2024} compared a constant-depth measurement-based method to a linear-depth unitary method for GHZ state generation on a $27$-qubit IBM quantum hardware. Both methods considered a linear qubit connectivity to generate the GHZ states.

The unitary method outperformed the measurement-based approach, which was partly attributed to inefficient implementation of classical feedforward on this IBM device. Moses et. al.~\cite{Moses2023} tested a log-depth unitary and a constant-depth measurement-based method on ion-trap devices, achieving high fidelities ($>0.5$) for up to $32$ qubits, and identified the measurement-based method as promising for devices with limited qubit connectivity. Liao et al.~\cite{Liao2025} proposed a unitary, BFS-based approach, which considers the qubit connectivity layout of the $156$-qubit IBM hardware, achieving multipartite entanglement in GHZ states of up to $75$ qubits using error suppression and sparse parity checks. Bao et al.~\cite{Bao_2024} enhanced unitary GHZ state generation by leveraging device connectivity and discrete time crystal techniques, demonstrating multipartite entanglement in GHZ states of up to $60$ qubits on a superconducting qubit device. Chen et al.~\cite{Kuan-Cheng} employed error mitigation techniques to reduce the circuit depth required for GHZ state preparation. GHZ states can also be generated via graph state transformations~\cite{graph_org, Dahlberg_2020}. For instance, a GHZ state was derived from a linear cluster state in~\cite{de_Jong_2024}, though the resulting GHZ state had only half the number of qubits as the original cluster state.

While all these works so far achieved impressive results in generating large GHZ states using various error suppression and error correction techniques, the available qubit connectivity on the devices has not been rigorously considered, especially in measurement-based approaches. Our work addresses this gap by exploring two connectivity-aware protocols for GHZ state generation: a measurement-based protocol (i.e. \textit{merging} protocol) and a unitary-based protocol (i.e. \textit{growing} protocol). 

The \textit{merging} protocol was initially developed for random graph architectures without accounting for implementation on circuit-based hardware \cite{chelluri2025}. We adapt this protocol for local hardware layouts by considering practical constraints such as gate scheduling and average degree of graphs. It constructs small GHZ states on star-like subgraphs and recursively fuses them using mid-circuit measurements, enabling significant parallelization and shallower circuit depth through strategic subgraph selection.

In contrast, the \textit{growing} protocol, introduced in~\cite{Liao2025} is a purely unitary approach that initiates from a central qubit and expands in a breadth-first manner by entangling neighboring qubits, always respecting hardware topology. This method eliminates measurement overhead and improves robustness against noise, though often at the cost of deeper circuits. We begin by implementing both protocols on real quantum hardware (IBM Eagle r3 chip) to evaluate the fidelity of the generated GHZ states. However, due to the high error rates of current NISQ devices, we also perform extensive simulations to explore performance in near-future scenarios. These simulations cover a range of hardware-relevant topologies, including IBM’s 127-qubit Eagle architecture \cite{IBM_Eagle}, rectangular lattices inspired by Google’s Willow layout~\cite{Willow}, and Erdős–Rényi random graphs~\cite{erdosrenyi1959,erdosrenyi1960}, representative of distributed quantum devices. We assess performance using multiple figures of merit, such as circuit depth, the number of two-qubit gates, measurements, and the final GHZ state fidelity. The choice between unitary-based and measurement-based protocols presents a trade-off that depends on the specific performance metrics and hardware capabilities. Unitary-based approaches are currently more compatible with near-term quantum devices, as they avoid mid-circuit measurements and classical feedforward, which remain error-prone on today's NISQ hardware. In contrast, measurement-based protocols, characterized by shallower circuit depth, are likely to become more effective as quantum technology advances, particularly in the context of fault-tolerant or more error-resilient architectures where measurements and feedforward operations are more reliable and efficient.

The paper is organized as follows. In Section~\ref{sec:method}, we discuss the two protocols, the \textit{merging }protocol in Section~\ref{sec:merging},and the \textit{growing} protocol in Section~\ref{section:growing}. We further discuss the benchmarking of the two protocols in Section~\ref{section:sim_details} based on different figures of merit. Specifically, the performance evaluations on hardware are discussed in Section~\ref{sec:fidelity}. Further simulations on near-future  hardware and random qubit layouts are provided in Sections~\ref{section:circuit_samp_hardware} and \ref{section:circuit_samp_random}, respectively. Finally, Section~\ref{section:summary} summarizes the main results and outlines potential directions for future work.

\section{Methods}\label{sec:method}

In this section, we discuss two protocols: the measurement-based merging protocol and the unitary-based growing protocol. We demonstrate both protocols on the IBM Eagle r3 chip layout; however, they are applicable to any architecture with arbitrary qubit connectivity.

\subsection{Merging Protocol}\label{sec:merging}

The main idea of this protocol is the ``merging" of multiple small GHZ states via measurements to create a large GHZ state efficiently. The protocol consists of the following four main steps.

\bigskip

\begin{figure*}[htbp!]
    \centering
    \includegraphics[width=0.7\linewidth]{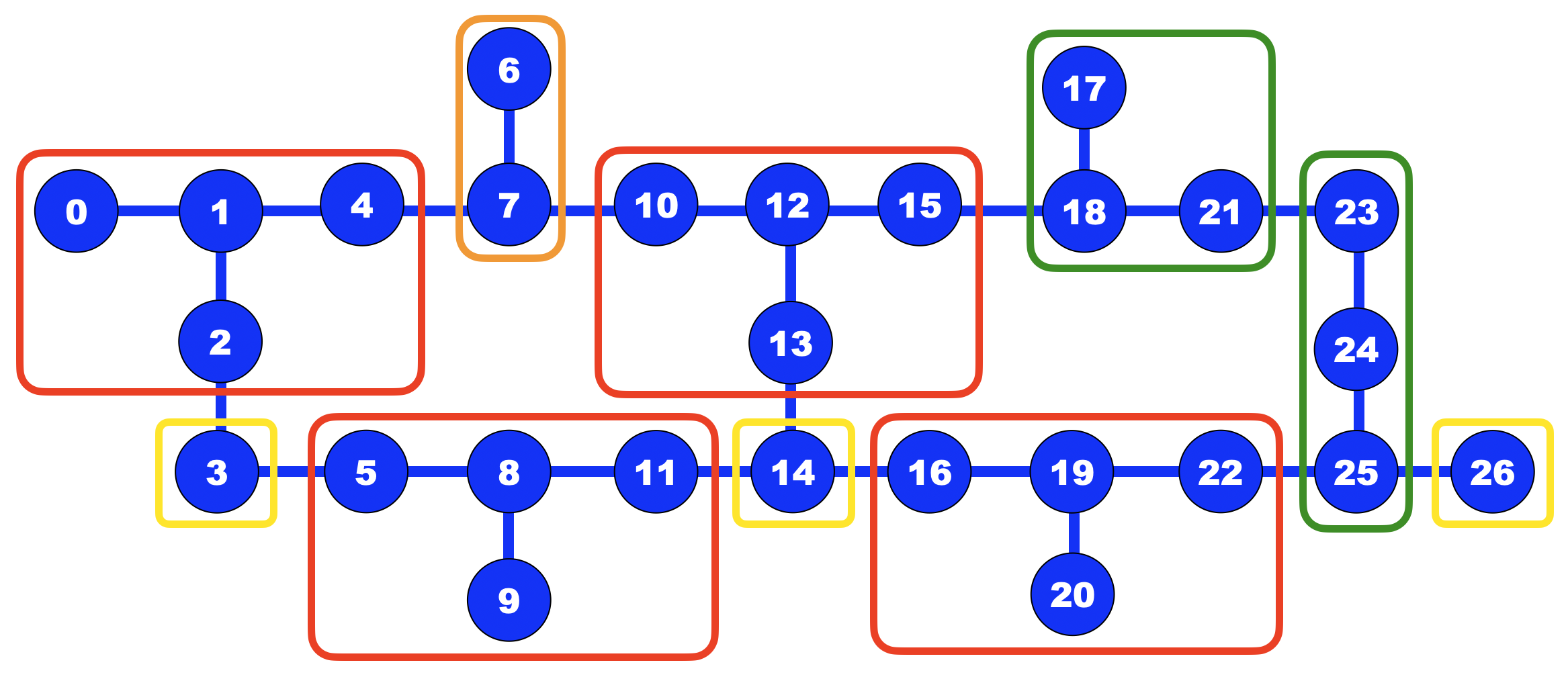}
    \caption{The figure illustrates star selection in the IBMQ device layout for the merging method. Coloured boxes indicate 4-, 3-, 2-, and 1-qubit stars (red, green, orange, and yellow, respectively), selected by iteratively choosing the highest-degree node. These stars, consisting of a central qubit and its neighboring leaves, form the basic units of the merging protocol.}
    \label{fig:ibm_layout}
\end{figure*}

\noindent\textbf{Step 1: star selection} \\
This step involves identifying groups of nodes, referred to as \textit{stars}, within the layout graph. First, we select the node with the highest degree in the graph layout, along with its neighboring nodes. This first \textit{star} is then removed from the layout, and the procedure is repeated: the next highest-degree node (among the remaining nodes) is selected along with its neighbors, forming the next star. This continues iteratively until all nodes have been grouped to a star. An illustrative example for star selection on the layout of an IBM device is shown in Figure~\ref{fig:ibm_layout}, where the selected stars are highlighted by colored boxes. The subgraphs in these boxes resemble stars.

\bigskip

\noindent\textbf{Step 2: construction of small GHZ states} \\
Each selected star is used to construct a GHZ state through a sequence of controlled-X ($CX$) gates. For the central node \( i \) with its neighboring nodes \(\mathcal{N}(i)\), the corresponding GHZ state is given by
\begin{equation}
    \ket{\mathrm{GHZ}}_{i,\mathcal{N}(i)} = \prod_{j \in \mathcal{N}(i)} CX_{i,j}\, H_i\, \ket{0}_{i,\mathcal{N}(i)},
\end{equation}
where \( H_i \) denotes the Hadamard gate applied to the central qubit \( i \), and \( CX_{i,j} \) represents the $CX$ gate with control qubit \( i \) and target qubit \( j \).

\bigskip

\noindent\textbf{Step 3: merging of GHZ states} \\
After constructing the small GHZ states, they are recursively merged to form a single large GHZ state. The protocol for merging two GHZ states involves applying a $CX$ gate followed by a single-qubit measurement. For example, consider two GHZ states, $\ket{\text{GHZ}_{n+1}}_{A_1B_{1:n}}$ and $\ket{\text{GHZ}_{m+1}}_{C_1D_{1:m}}$, of sizes $n+1$ and $m+1$, respectively. These states involve the qubits $A_1$ and $B_{1:n} = \{B_1, B_2, \dots, B_n\}$, and $C_1$ and $D_{1:m} = \{D_1, D_2, \dots, D_m\}$. If qubits $A_1$ and $C_1$ share an edge, the two GHZ states can be merged into a new GHZ state of size $n + m + 1$ by applying a controlled-NOT gate $CX_{A_1, C_1}$, followed by measuring qubit $C_1$ in the $X$-basis. Depending on the measurement outcome $x \in \{0,1\}$, a corrective $X$ gate may need to be applied to the qubits $D_{1:m}$ to obtain the desired GHZ state exactly. A more detailed discussion of this merging procedure is provided in Appendix~\ref{appendix_heurstic_protocol}.

\bigskip

\noindent\textbf{Step 4: re-adding/re-using the measured qubits} \\
During each merge, a qubit is measured (see Step~3). Afterwards, this qubit is re-integrated into the GHZ state by initializing it to \(\ket{0}\) and applying a \(CX\) gate between the measured qubit and one of its neighboring qubits already in the GHZ state.

\bigskip

Note that in step 1, it is also possible to select stars based on alternative criteria rather than choosing nodes with the highest degree. We explore this option extensively in Appendix~\ref{app:eval_star_selec_crit}, where stars are chosen such that the degree of each star is the same or nearly the same (throughout the article, degree of star/star degree refers to the degree of the star’s center node), resulting in similar star sizes.

The motivation behind this approach is to reduce the total time required for constructing all stars in step 2. Since the $CX$ gates involved in this step are applied sequentially and their number depends on the degree of the star, the total time to prepare the state scales with the star size. The preparation time of a star is considered in our benchmark in the circuit depth. A shorter preparation time is indicated by a smaller circuit depth. By choosing similar star sizes, we aim to reduce the circuit depth of the star preparation. However, this may come at the cost of requiring more two-qubit gates and measurements overall, due to an increasing number of stars and thus an increasing number of merges. To implement this, we consider two possibilities: First, we define a scaling factor as the ratio of the chosen star degree to the average degree of the layout graph. A scaling factor of exactly 1 indicates that the selected star degree matches the average degree. If the factor is less than 1, the stars degrees are smaller than the average degree; if it is greater than 1, the stars degrees are larger than the average degree. Secondly, we also set the star size explicitly to fixed values. Note that it is not always possible to select all stars with exactly the same size, as the initial layout graph is fixed. In such cases, stars are chosen with the closest possible size to the target value.

\subsection{Growing Protocol}\label{section:growing}

In addition to the \textit{merging} approach, we present here a purely unitary method for efficiently generating a GHZ state, tailored to a given qubit connectivity layout. The growing method is explained in the stepwise fashion as follows:\\

\begin{figure*}[t]
    \centering
    \begin{tabular}{ccc}
    \includegraphics[width=0.3\textwidth]{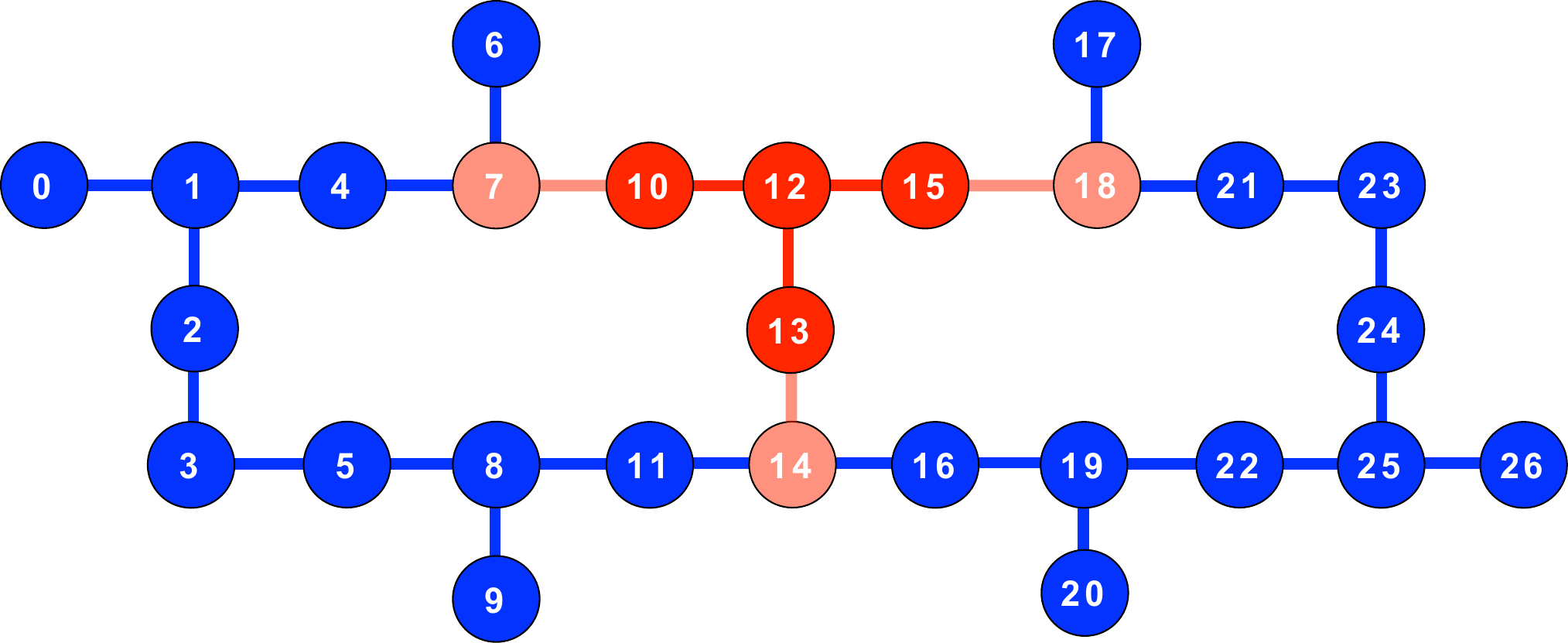}&
    \hfill
    \includegraphics[width=0.3\textwidth]{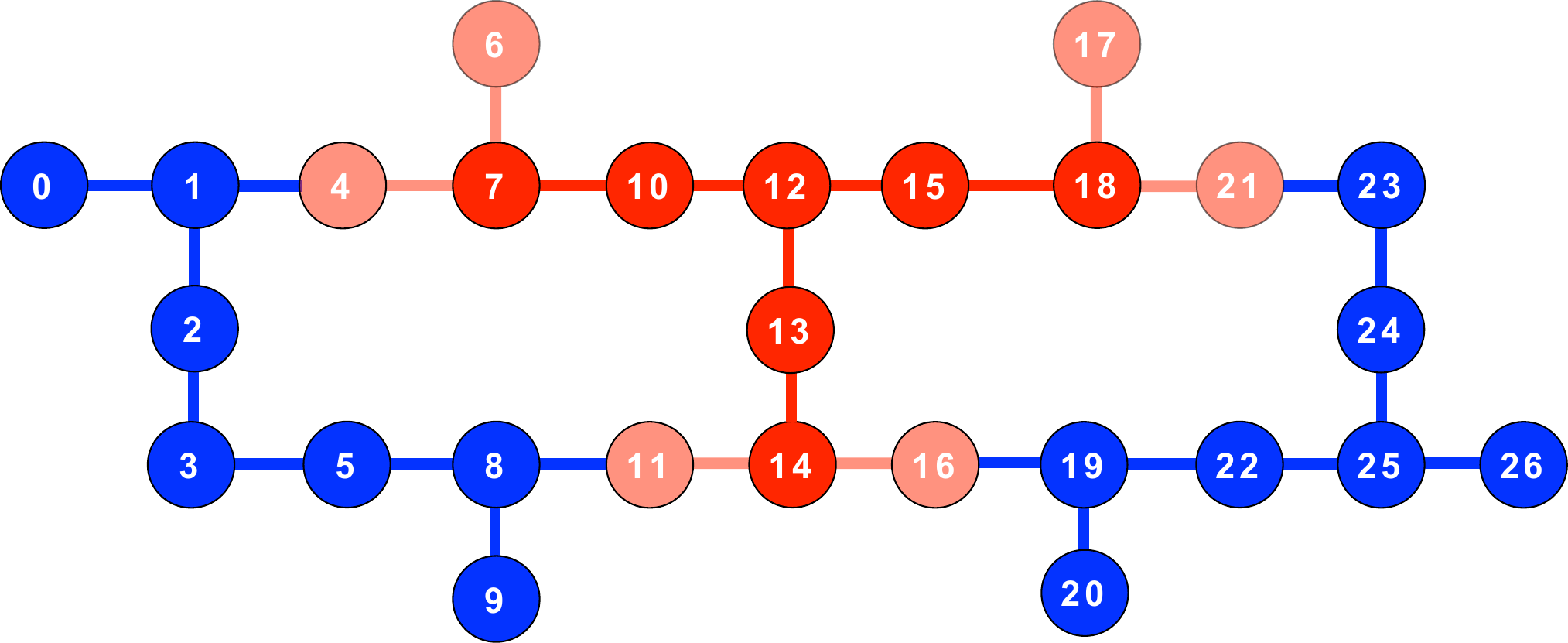}&
    \hfill
    \includegraphics[width=0.3\textwidth]{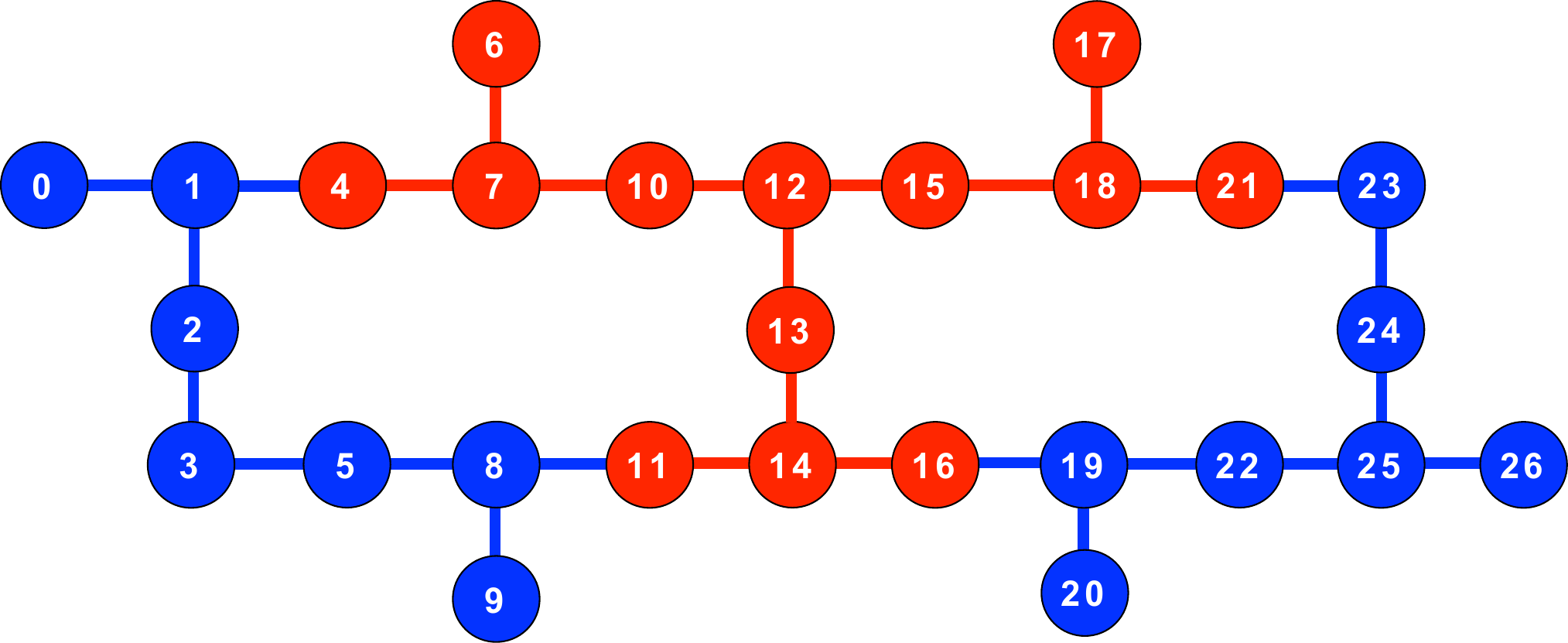}\\
    (a) & (b) & (c)\\
     \end{tabular}
    \caption{The first three steps of the GHZ state growing method on the IBM layout as shown in this figure. The red nodes correspond to qubits that are included in the current GHZ state and the light red nodes are to-be-included in the next step. Panel (a) shows the initial GHZ state generation according to Equation~\eqref{eq:init_ghz_state}, picking node $12$ as the highest degree node and the first BFS step identifying the direct neighbors of the initial GHZ state. Panel (b) shows the GHZ state after adding the neighbors from the first BFS step via $CX$ gates and the second BFS step, identifying the direct neighbors of the current GHZ state. Panel (c) shows the final GHZ state after adding the neighbors from the second BFS step, reaching a final size of $13$ qubits.}
    \label{fig:ghz_state_growing_method}
\end{figure*}

\noindent \textbf{Step 1: initial GHZ state generation}\\
We begin by selecting the node $i$ with the highest degree and construct an initial GHZ state involving this node and its immediate neighbors $\mathcal{N}(i)$ (cf. Figure~\ref{fig:ghz_state_growing_method} (a)):
\begin{equation}
    \ket{\mathrm{GHZ}}_{i, \mathcal{N}(i)} = \prod_{j \in \mathcal{N}(i)} CX_{i,j} H_i \ket{0}_{i, \mathcal{N}(i)},
    \label{eq:init_ghz_state}
\end{equation}
where $H_i$ is the Hadamard gate acting on the qubit corresponding to the highest degree node $i$ in the layout graph and $CX_{i,j}$ is the controlled-$X$ gate with control qubit $i$ and target qubit $j$. \\

\noindent\textbf{Step 2: expanding the GHZ state via BFS}\\
Next, we expand the state by identifying the neighbors of the qubits already in the GHZ state and include them via $CX$ gates (cf. Figure~\ref{fig:ghz_state_growing_method} (b) and (c)). Therefore, the qubits already in the GHZ state have to be the control qubits of the $CX$ gates and the neighboring qubits have to be the target qubits, such that the resulting state is again a GHZ state. This step is repeated until the desired state size is reached.\\

While this method avoids any merging operations and relies solely on unitary gates, the BFS naturally incorporates the qubit connectivity of the underlying hardware. Although we anticipate that this approach may yield deeper circuits than our \textit{merging} protocol, it is expected to outperform other purely unitary techniques, such as linear constructions~\cite{Baumer2024} or logarithmic-depth methods that ignore qubit connectivity~\cite{Moses2023}-in terms of overall circuit depth.  

Because this approach begins with a small GHZ state centered on the highest-degree node and incrementally grows it through BFS, we refer to it as GHZ state \textit{growing} for the rest of the article. We will compare the proposed \textit{merging} protocol to this unitary method in terms of the resulting circuit depth, measurement overhead, number of two qubit gates and for the IBM layout also in terms of the resulting GHZ state fidelity.

\section{Benchmark Results} \label{section:sim_details}

 To benchmark the two methods presented in Sections~\ref{sec:merging} and~\ref{section:growing}, we implemented them on IBM’s 127-qubit Eagle chip topology \cite{IBM_Eagle}, using fidelity as the primary figure of merit. The results were compared with noisy simulations of the same hardware, as discussed in Section~\ref{sec:fidelity}. While running experiments on real quantum hardware is both interesting and valuable, the current devices are still subject to significant noise and errors. As a result, such experiments alone are insufficient for estimating the feasibility of these protocols on near-future devices. To address this, we conducted additional simulations across several graph layouts: IBM’s 127-qubit Eagle topology \cite{IBM_Eagle}, rectangular grids inspired by Google’s Willow architecture \cite{Willow} (both detailed in Section~\ref{section:circuit_samp_hardware}), and randomly generated graphs representing distributed quantum computing scenarios (discussed in Section~\ref{section:circuit_samp_random}). For these simulations, we evaluated several figures of merit, including circuit depth, the number of two-qubit gates, and the number of measurements across all layouts. The circuit depth is measured before hardware-specific compilation and is thus independent of any backend. Single-qubit gates alone are not explicitly discussed, as their error rates are typically an order of magnitude lower than those of two-qubit gates. Nevertheless, they are included in our circuit-depth calculations. All simulations were implemented in Python using the \texttt{NetworkX} \cite{networkx} and \texttt{Qiskit} \cite{qiskit2024} libraries. Each configuration was executed 100 times to gather statistical data.

\subsection{Fidelity analysis for IBM Eagle chip}\label{sec:fidelity}

Quantum state tomography can fully characterize any quantum state. However, it quickly becomes infeasible as the number of measurements grows exponentially with the number of qubits. To benchmark the performance of our protocols, we therefore rely on the Hellinger fidelity~\cite{hellinger_fidelity_qiskit, Hellinger_1909} as our evaluation metric. It quantifies the similarity between two (discrete) probability distributions $P$ and $Q$ (e.g., ideal and measured outcomes) as:
\begin{equation}
\label{eq:hellinger_fidelity}
\begin{split}
    F_H(P, Q) &= \left[1 - \left(\frac{1}{\sqrt{2}} \left\| \sqrt{P} - \sqrt{Q} \right\|_2\right)^2\right]^2\\
&=\left(\sum_i\sqrt{p_iq_i}\right)^2
\end{split}
\end{equation}

where \( p_i \) and \( q_i \) are the probabilities of \( P \) and \( Q \) respectively, and \( \left\| \cdot \right\|_2 \) denotes the Euclidean norm. \\

\begin{figure*}
    \begin{tabular}{cc}
    \includegraphics[width=0.48\textwidth]{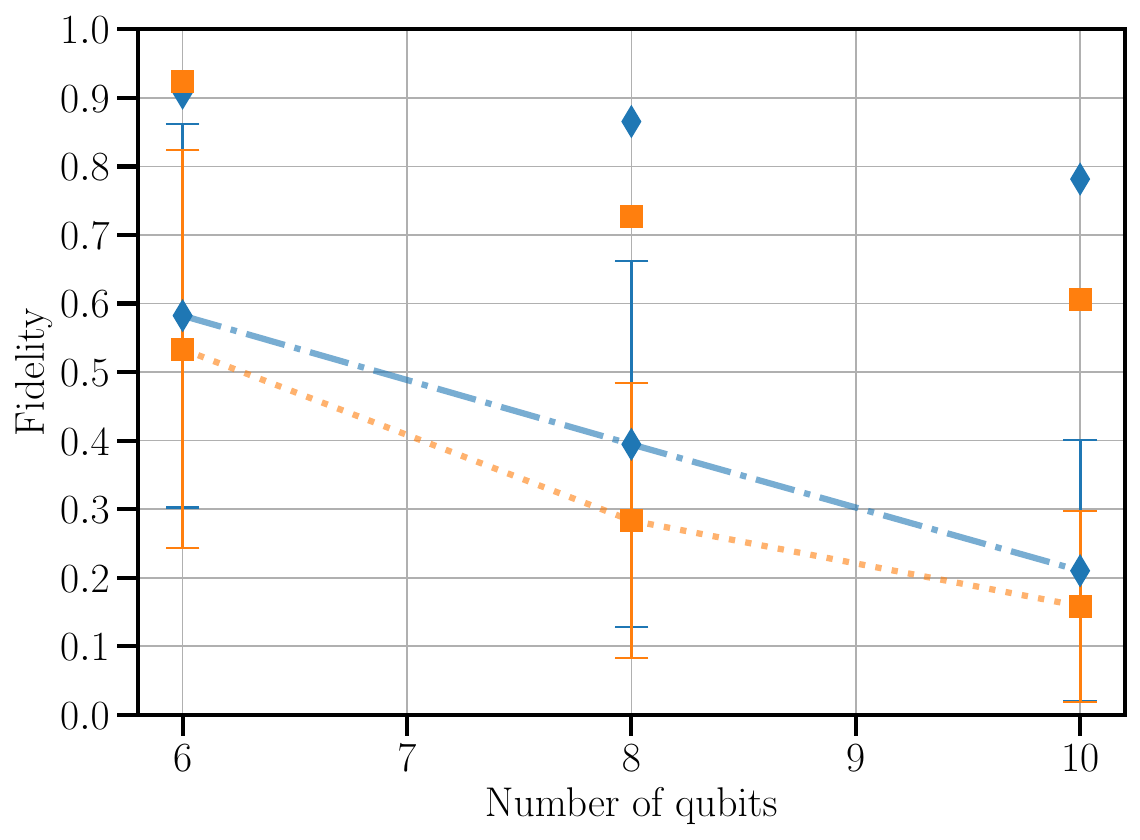}&
    \includegraphics[width=0.48\textwidth]{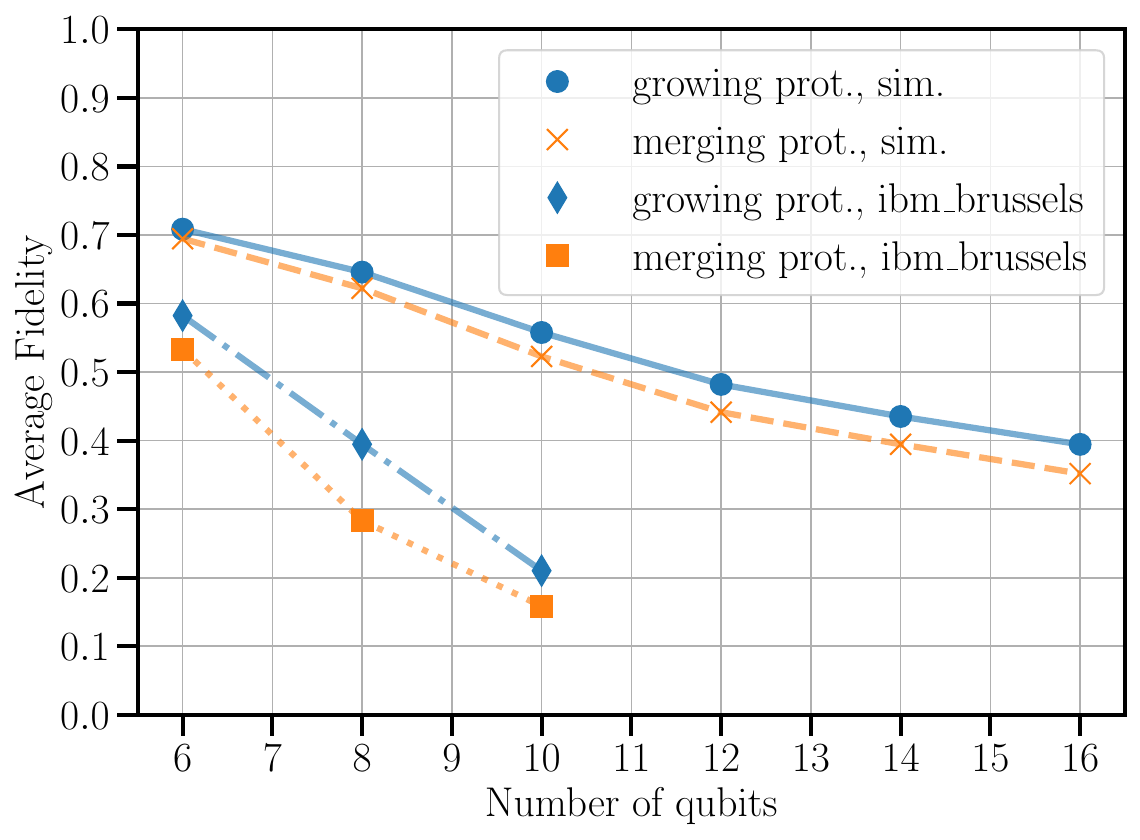}\\
    (a) & (b)\\
    \end{tabular}
    \caption{GHZ state fidelity (cf. Equation~\eqref{eq:hellinger_fidelity}) over the number of qubits in the final GHZ state. Here we compare the \textit{growing} protocol (blue circles, blue diamonds) with our \textit{merging} protocol (orange crosses, orange squares) using the highest degree star size. For each GHZ state size and each protocol, we created and executed $100$ different circuits. Panel (a) shows the average fidelity values obtained from executing the circuits of the \textit{growing} protocol (blue diamonds) and the \textit{merging} protocol (orange squares) on quantum hardware \texttt{IBM\_Brussels} and additionally the maximum fidelity achieved for each size is indicated as another marker. The standard deviation is shown as error bars on the average values. Panel (b) shows the average fidelity values obtained from noisy simulations using the full \texttt{IBM\_Brussels} noise model (blue circles, orange crosses) and from running the same circuits on the quantum hardware \texttt{IBM\_Brussels} (blue diamonds, orange squares). The legend of Panel (b) is also valid for Panel (a). For each GHZ state size, we created and executed $100$ different circuits. Each circuit as been executed with $4096$ shots.}
    \label{fig:mean_fidelity_full_noise+hardware}
\end{figure*}

We use the Hellinger fidelity (Equation~\ref{eq:hellinger_fidelity}) to evaluate the performance of our two protocols. To this end, we implemented GHZ state generation circuits on the 127-qubit Eagle-chip device (\texttt{IBM\_Brussels}). For each GHZ state size and each protocol, we generated and executed 100 random circuits on \texttt{IBM\_Brussels}, each using a different set of physical qubits, and computed the average fidelity across these runs. These averages are presented in Figure~\ref{fig:mean_fidelity_full_noise+hardware} (a), along with the highest fidelity observed among the 100 instances, illustrating the best-case performance. The standard deviations, shown as error bars on the average values, indicate substantial variation in fidelity for a given state size. This spread is largely due to heterogeneous two-qubit gate and measurement error rates across the device. Notably, some instances yield high-fidelity GHZ states despite overall variability. To complement the hardware experiments, we performed noisy simulations using the full noise model of the same device. The simulated fidelities are shown in Figure~\ref{fig:mean_fidelity_full_noise+hardware} (b). While the simulations rely on a noise model calibrated from the hardware itself, and simulations and experiments were conducted within the same calibration window, we observe a consistent discrepancy between simulated and experimental results.

Furthermore, we find that the \textit{growing} protocol outperforms the \textit{merging} protocol in terms of fidelity. This is expected, as the \textit{growing} protocol avoids mid-circuit measurements and also requires slightly fewer two-qubit gates, both of which contribute significantly to noise accumulation. It should be noted that no error mitigation techniques were employed to enhance the fidelity in our results.

\subsection{Performance on quantum hardware layouts} \label{section:circuit_samp_hardware}

\begin{figure*}
\begin{tabular}{ccc} 
    \includegraphics[width=0.3\textwidth]{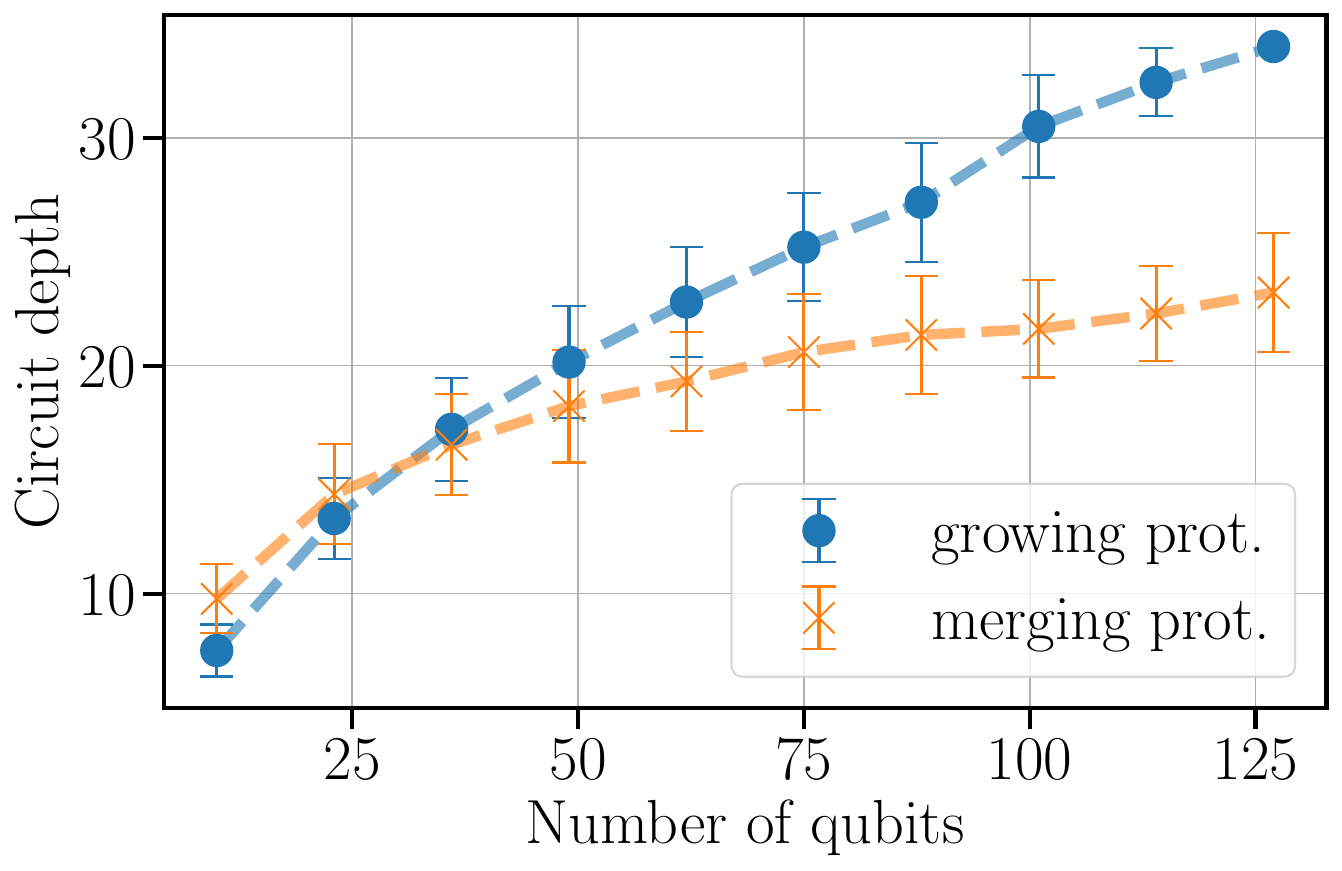}&
    \includegraphics[width=0.3\textwidth]{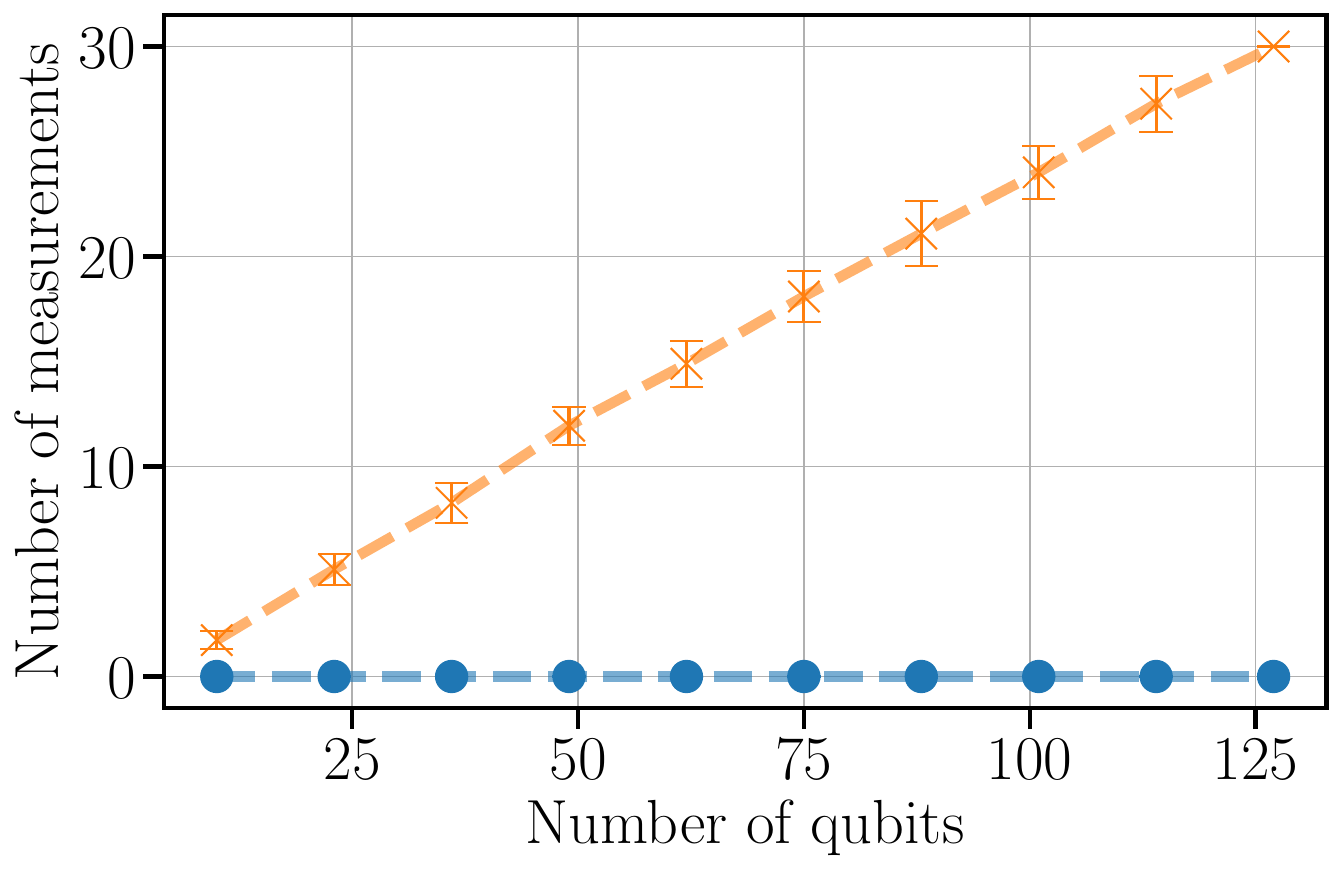}&
    \includegraphics[width=0.3\textwidth]{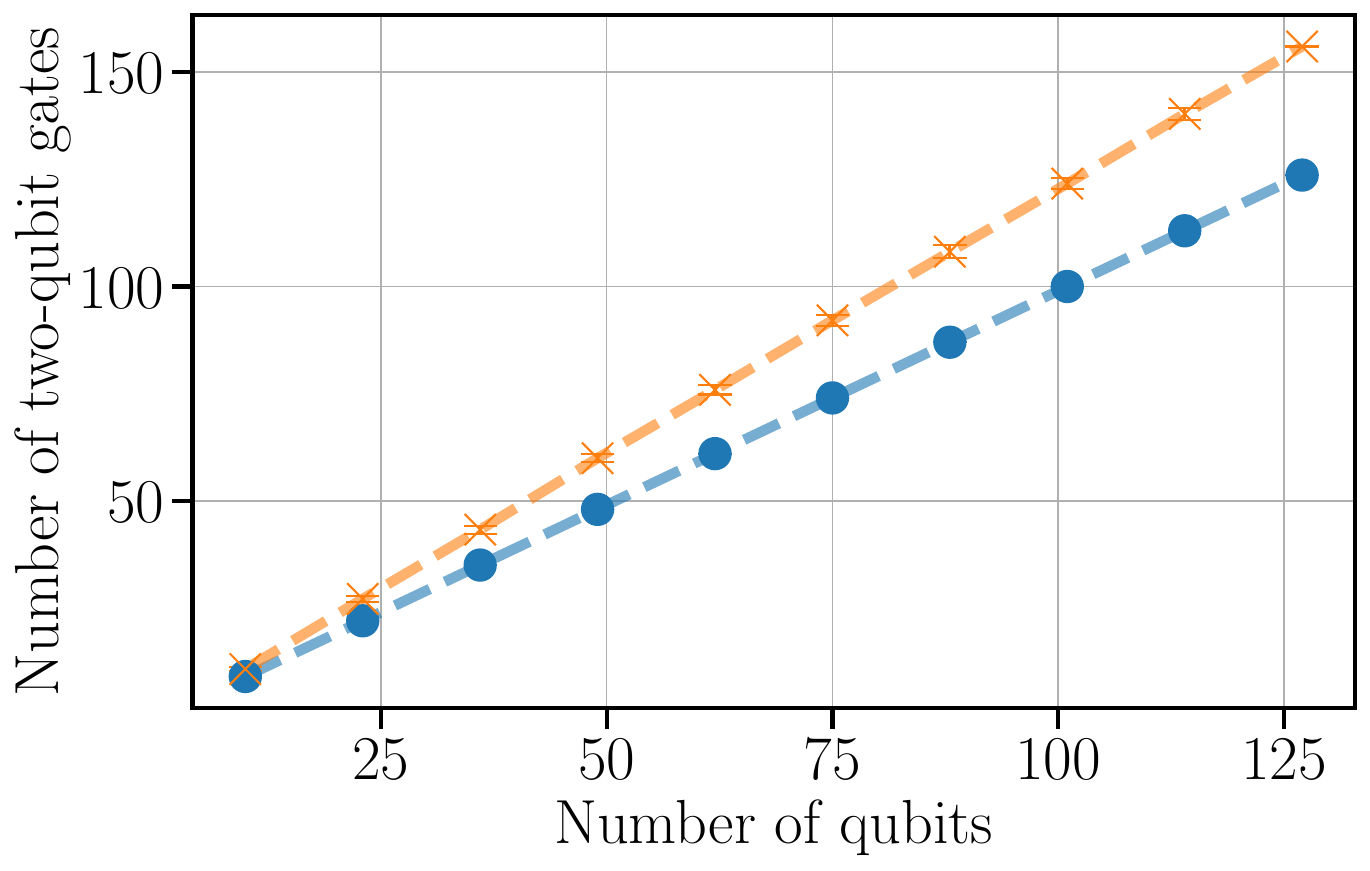}\\
    (a) & (b) & (c) \\
    \end{tabular}
    \caption{Random subgraph sampling from IBM's 127-qubit Eagle layout \cite{IBM_Eagle}. Here we compare the \textit{growing} protocol (blue circles) with our \textit{merging} protocol (orange crosses) using the highest degree star size. Panels (a), (b) and (c) show the averaged circuit depth, number of measurements and number of two-qubit gates in dependence of the number of qubits in the final GHZ state. For each GHZ state size, we generated $100$ subgraph samples of this size randomly from the initial layout graph. The error bars show the standard deviation obtained from averaging over the samples.}
    \label{fig:ibm_brisbane_merge_vs_grow}
\end{figure*}

\begin{figure*}
\begin{tabular}{ccc}

    \includegraphics[width=0.3\textwidth]{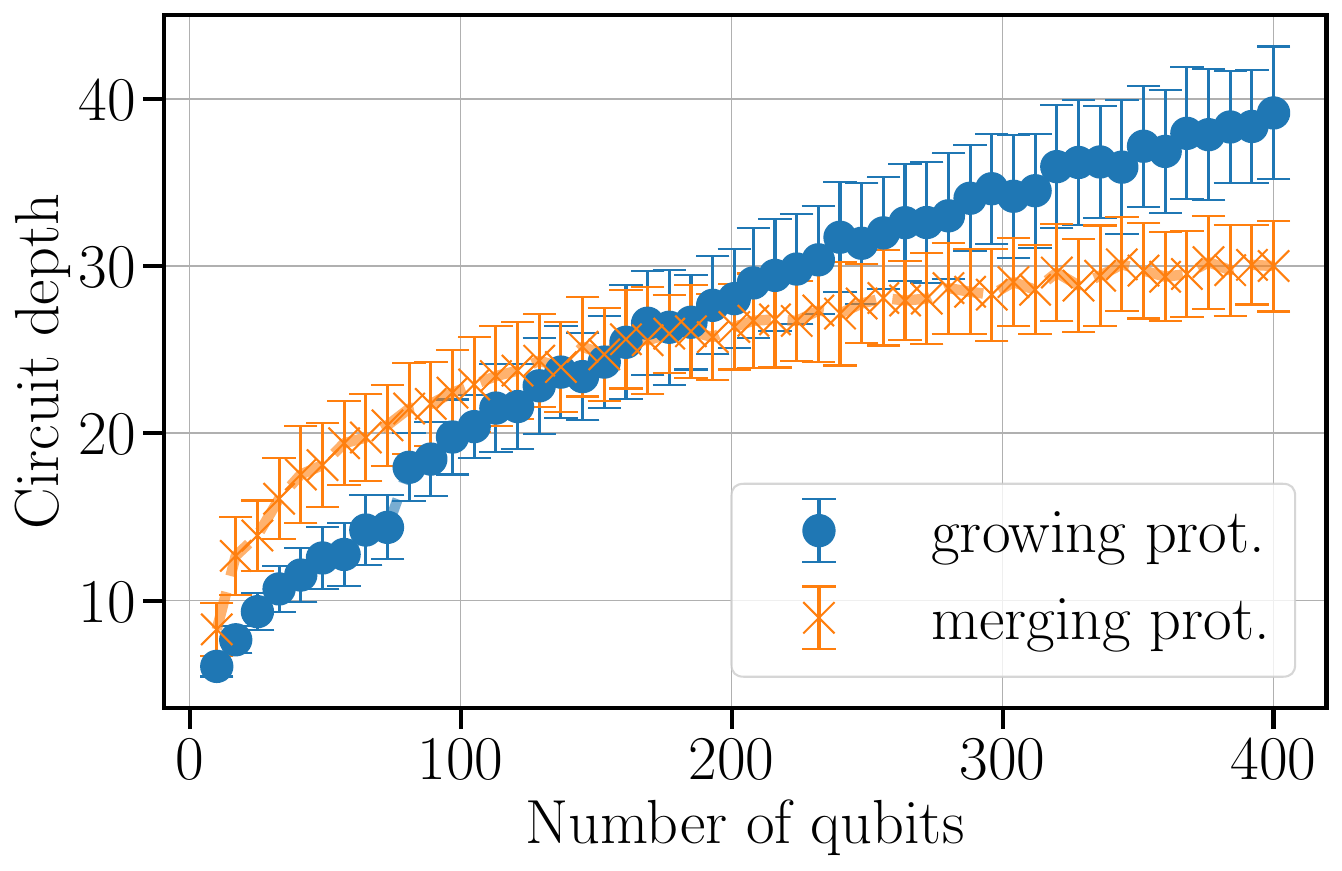}&
    \includegraphics[width=0.3\textwidth]{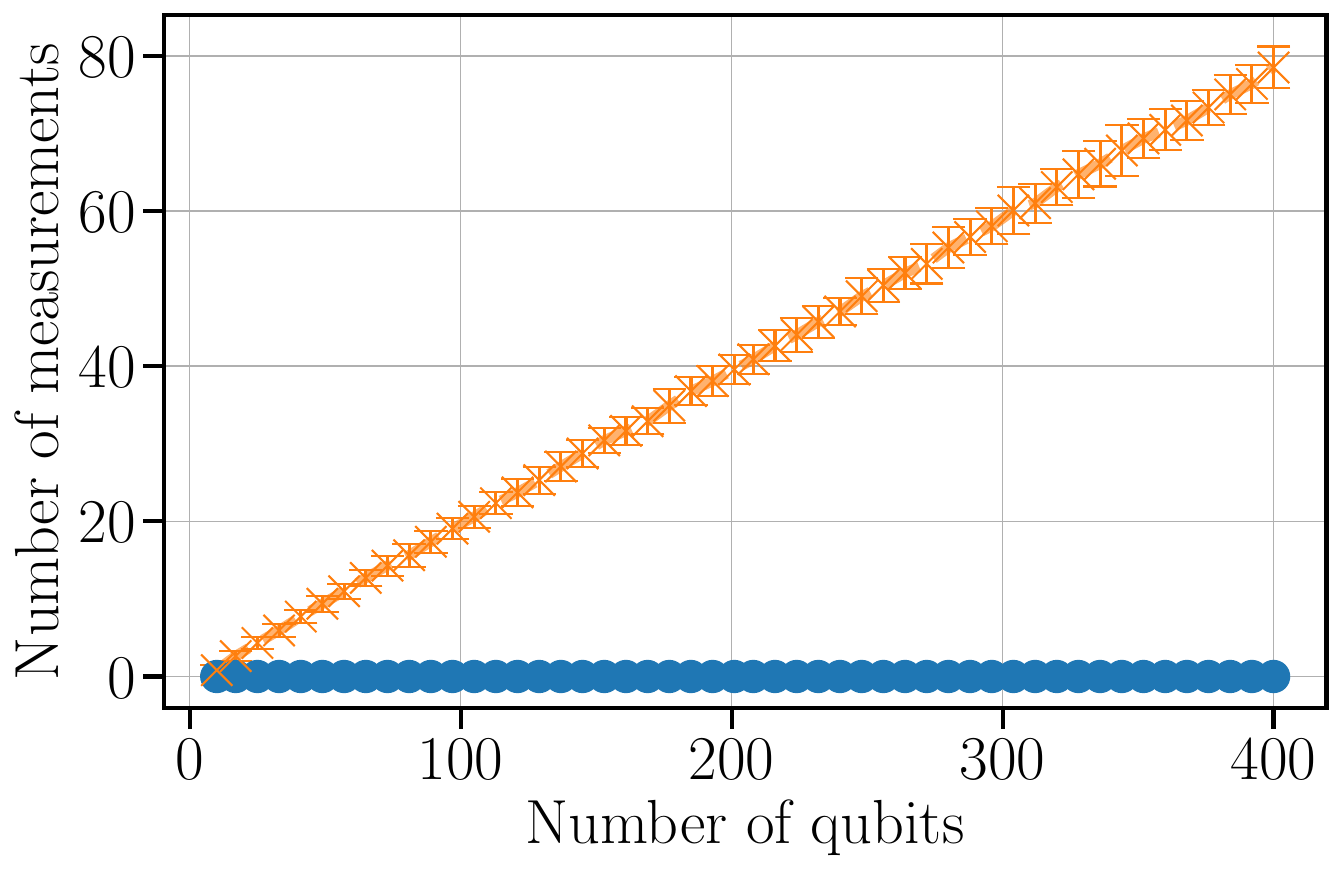}&
    \includegraphics[width=0.3\textwidth]{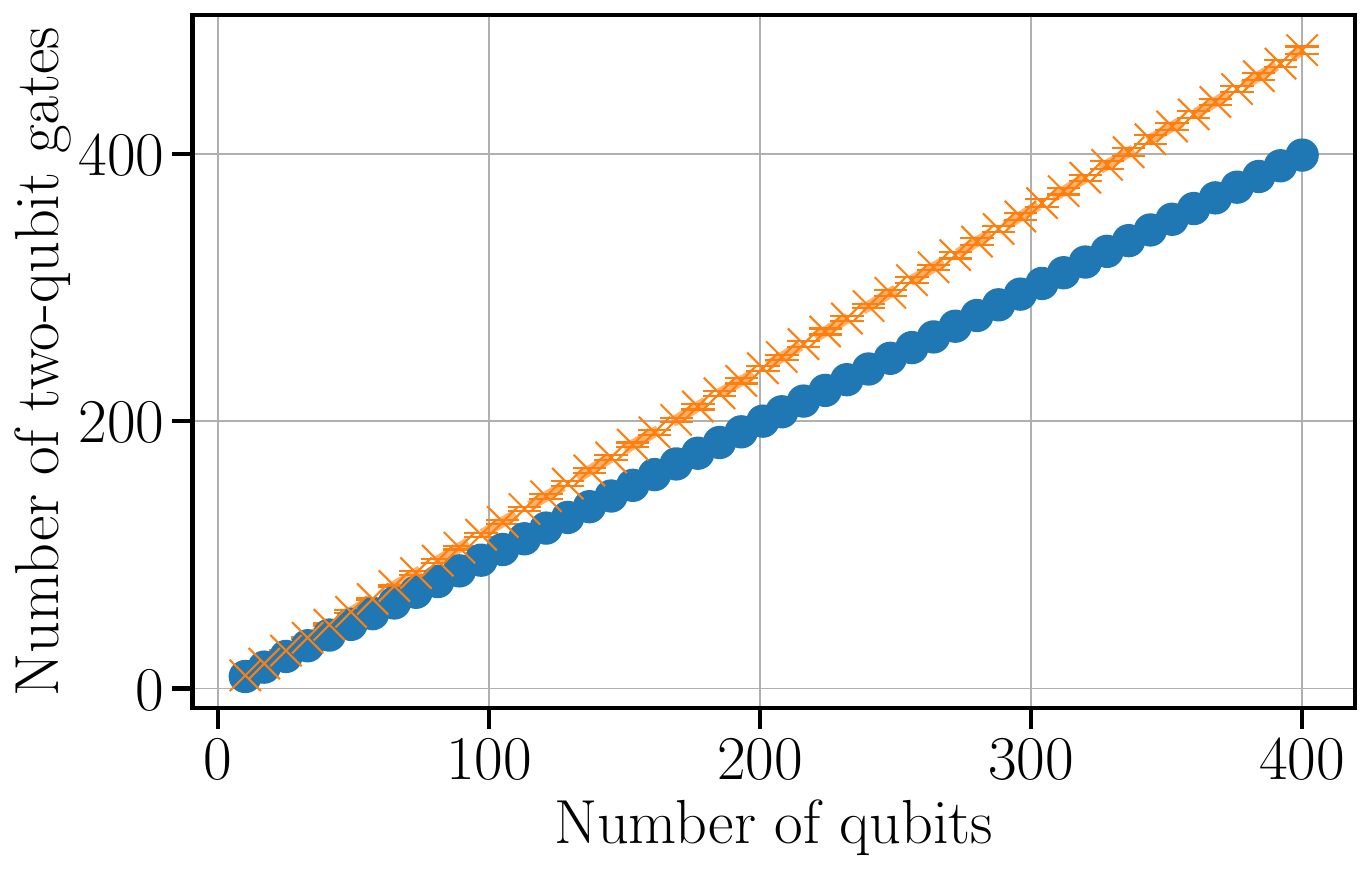}\\
    (a) & (b) & (c) \\
    \end{tabular}

    \caption{Random subgraph sampling from rectangular grid layout inspired by Google's Willow chip \cite{Willow}. Here we compare the \textit{growing} protocol (blue circles) with our \textit{merging} protocol (orange crosses) using the highest degree star size. Panels (a), (b) and (c) show the averaged circuit depth, number of measurements and number of two-qubit gates in dependence of the number of qubits in the final GHZ state. For each GHZ state size, we generated $100$ subgraph samples of this size randomly from the initial layout graph. The error bars show the standard deviation obtained from averaging over the samples.}
    \label{fig:rect_grid_merge_vs_grow}
\end{figure*}

In this section, we examine the results for IBM's 127-qubit Eagle layout \cite{IBM_Eagle} and Google's Willow \cite{Willow} layout shown in Figures \ref{fig:ibm_brisbane_merge_vs_grow} and \ref{fig:rect_grid_merge_vs_grow} respectively. We compare the \textit{merging} protocol with the \textit{growing} protocol on grounds of averaged circuit depth, number of measurements and two-qubit gates.

Interestingly, we observe that the \textit{merging} protocol performs best when selecting stars based on the highest degree. Consequently, only the highest-degree \textit{merging} protocol is compared with the \textit{growing} protocol, while the comparison of different star selection strategies within the \textit{merging} protocol is discussed in Appendix~\ref{app:eval_star_selec_crit}. As shown in Figures~\ref{fig:ibm_brisbane_merge_vs_grow} and~\ref{fig:rect_grid_merge_vs_grow}, the \textit{merging} protocol results in shallower circuits compared to \textit{growing}, with the improvement becoming more pronounced as the GHZ state size increases. However, this advantage comes with a trade-off: unlike \textit{growing}, \textit{merging} requires measurements and leads to a higher number of two-qubit gates due to qubit reuse.

\begin{figure*}[tbh!]
\begin{tabular}{ccc}

    \includegraphics[width=0.3\textwidth]{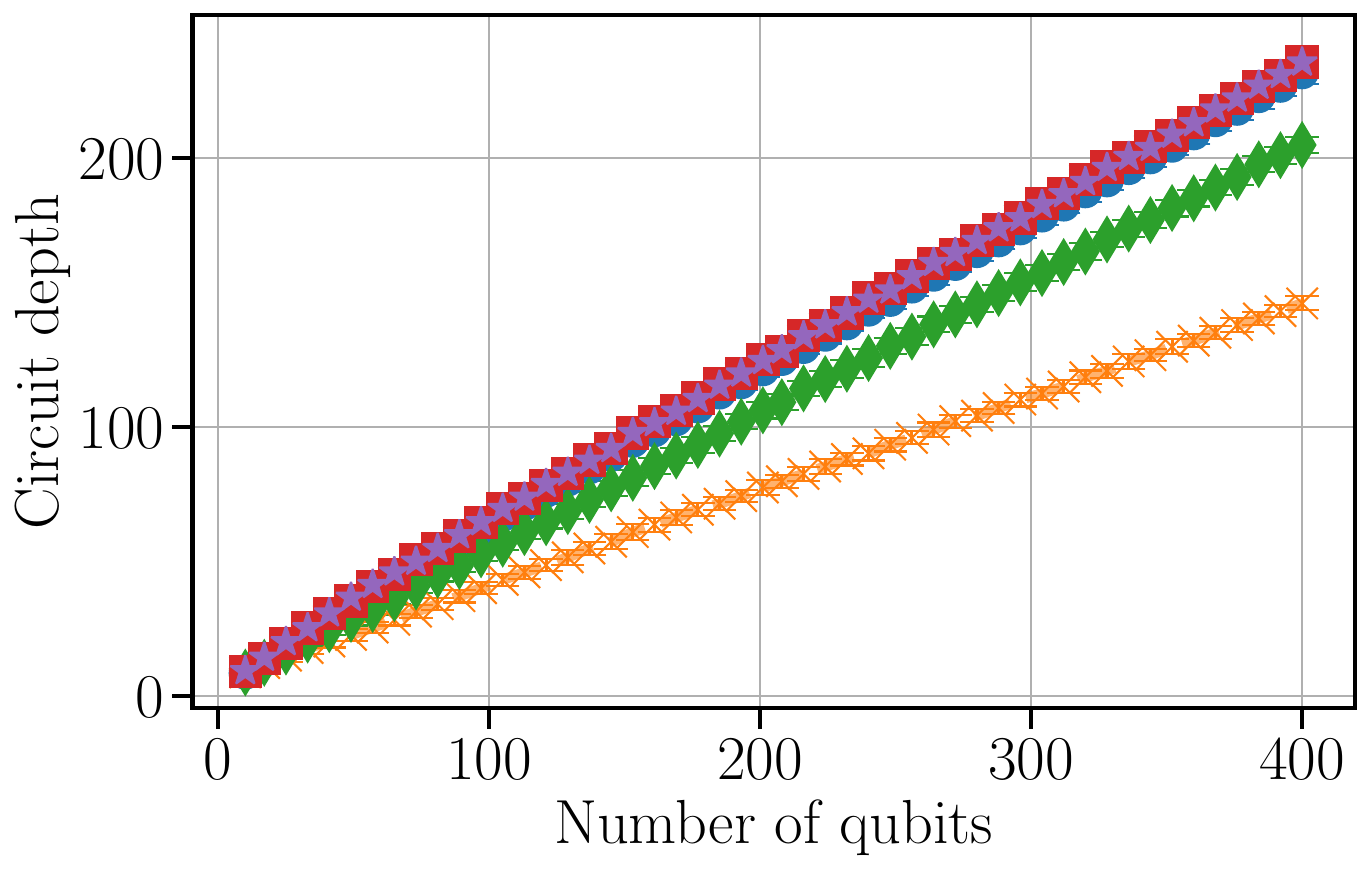}&
    \includegraphics[width=0.3\textwidth]{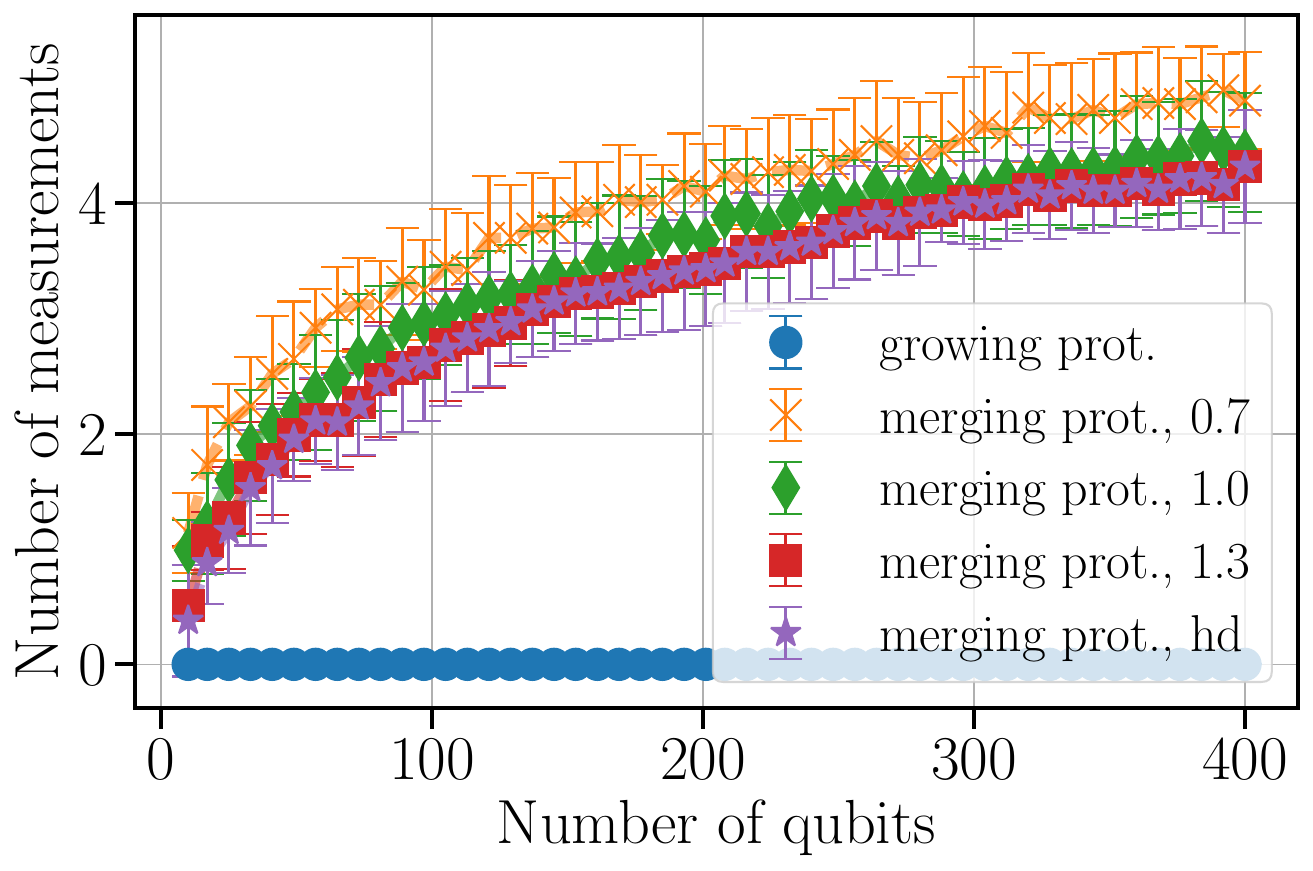}&
    \includegraphics[width=0.3\textwidth]{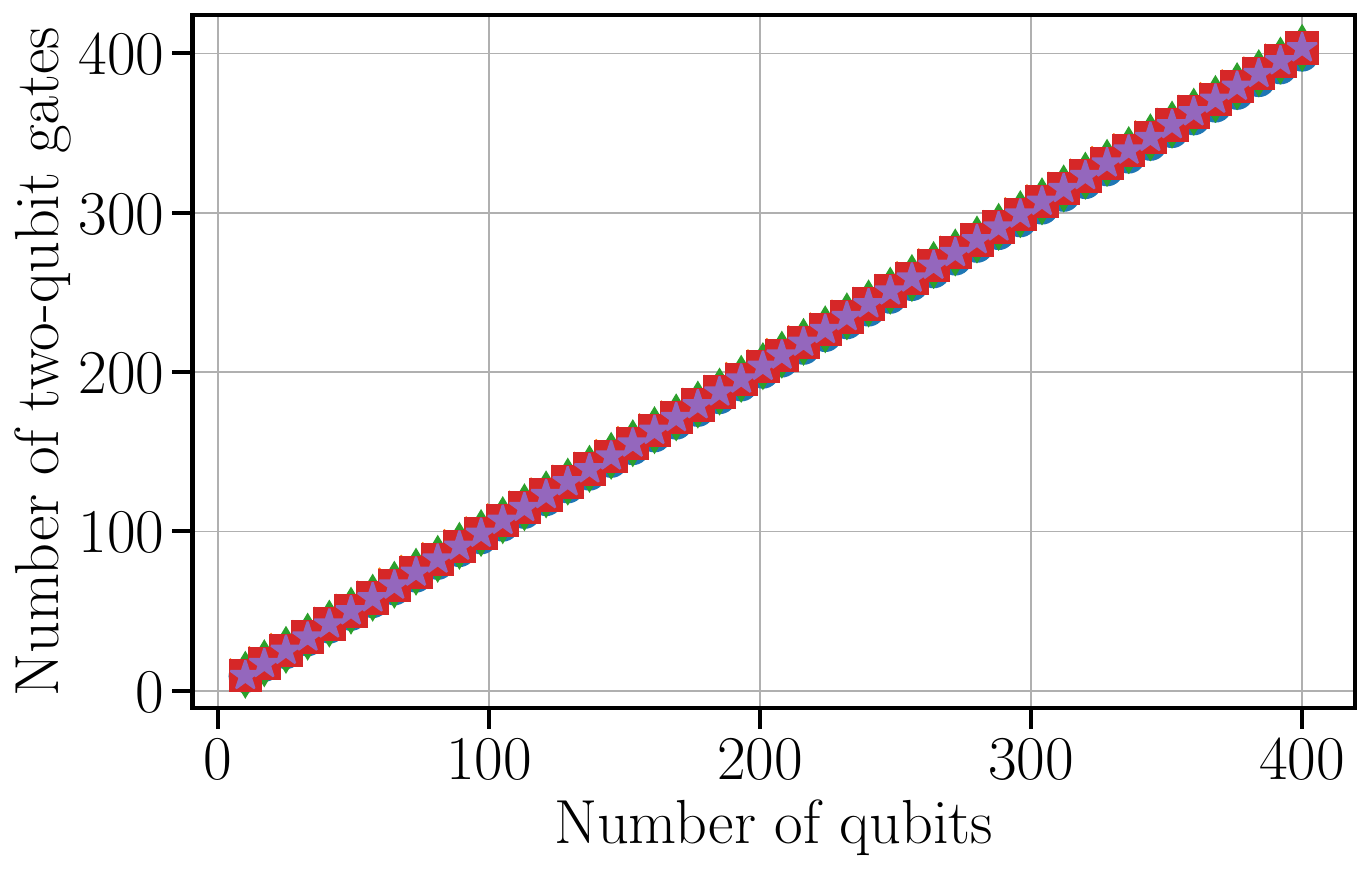}\\
     (a) & (b) & (c) \\
\end{tabular}

   \caption{Sampling from Erd\H{o}s--R\'enyi random graphs~\cite{erdosrenyi1959,erdosrenyi1960}. Here we compare the \textit{growing} protocol with our \textit{merging} protocol using a varying target size for the star selection. The curves represent \textit{growing} protocol (blue circles) as well as the \textit{merging} protocol with a star selection based on scaling factors $0.7$ (orange crosses), $1.0$ (green diamonds), $1.3$ (red squares) and the highest degree (violet stars). As explained in Section~\ref{sec:merging}, the scaling factor relates the target star size for every star selection during \textit{merging} with the average star size in the input graph. A larger scaling factor thus corresponds to a larger star size (cf. Panels (c), (f) and (i) of Figure~\ref{fig:random_graph_merge_vs_grow_num_qubits} in Appendix~\ref{app:eval_star_selec_crit}). For the graph generation in the Erd\H{o}s--R\'enyi model, we used $p=0.5$. Panels (a), (b) and (c) show the averaged circuit depth, number of measurements and number of two-qubit gates for \textit{merging} in dependence of the number of qubits in the final GHZ state. For each GHZ state size, we generated $100$ Erd\H{o}s--R\'enyi random graphs of this size. The error bars show the standard deviation obtained from averaging over those samples.}
    \label{fig:ER_graph_merge_vs_grow_pval}
\end{figure*}

\subsection{Performance on random graphs}\label{section:circuit_samp_random}

Random graph layouts naturally arise in modular quantum computing architectures, where small-scale quantum processors are interconnected to form a larger system. Examples include distributed quantum computing based on ion-trap architectures or optical fiber networks linking independent quantum nodes \cite{dqc_survey, dqc_ion}. In such settings, the physical connectivity between modules can be modeled as a graph structure. To model such scenarios, we simulated random graphs based on the Erd\H{o}s-R\'enyi model~\cite{erdosrenyi1959,erdosrenyi1960}, denoted as \( G(N, p) \), where \( N \) is the number of nodes and \( p \) is a fixed probability that any given pair of nodes is connected by an edge, independently of all other pairs. The Erd\H{o}s-R\'enyi model does not guarantee connectedness for arbitrary parameters. However, entanglement generation using LOCC (Local Operations and Classical Communication) requires the overall graph to be connected. Therefore, we adopted a modified version of the Erd\H{o}s-R\'enyi model introduced in~\cite{chelluri2025}, where connectivity is explicitly enforced. In this version, each node is initially connected to at least one other node to ensure connectivity, and additional edges are then added according to the standard \( G(N, p) \) process.

We studied the performance of the \textit{growing} and \textit{merging} protocols for creating GHZ states. We determined the circuit depth, number of measurements and the number of two-qubit gates for each generated circuit. For every GHZ state size we created $100$ circuits based on different random graphs. The averaged circuit depth, number of measurements and number of two-qubit gates are shown in Figure \ref{fig:ER_graph_merge_vs_grow_pval}  for the different GHZ state sizes. Unlike the results for the IBM or rectangular layout, we observe that selecting the stars with the highest degree is no longer the global best strategy. Therefore, we also consider another selection criteria which allows us to fix the target star size during the selection (see last paragraph of Section~\ref{sec:merging}). We adjust this target size in relation to the average star size in the input graph via a scaling factor, where larger scaling factors correspond to larger target sizes (cf. Panels (c), (f) and (i) in Figure~\ref{fig:random_graph_merge_vs_grow_num_qubits} in Appendix ~\ref{app:eval_star_selec_crit}).

We find that smaller star sizes lead to reduced circuit depth, while larger star sizes require fewer measurements during the \textit{merging} process. This reveals an inherent trade-off, where the optimal strategy must be chosen on a case-by-case basis. Additionally, in Appendix~\ref{app:eval_star_selec_crit}, we examine Erdős–Rényi graphs with varying values of $p$. Where we observe an overall increasing number of measurements with decreasing $p$ values. Due to the re-adding of qubits after merging (cf. step 4 in Section~\ref{sec:merging}), we get an additional two-qubit gate for each merge, i.e., for each measurement. Since the number of measurements is very small compared to the minimal required number of two-qubit gates to generate the GHZ state of large sizes, we see all five curves overlapping for the number of two-qubit gates in Panel (c) of Figure~\ref{fig:ER_graph_merge_vs_grow_pval}.

\section{Summary and outlook}\label{section:summary}

As quantum computers continue to improve, generating large-scale entangled states has become an important benchmarking tool. In this work, we explore the feasibility of creating GHZ states on quantum hardware using two distinct approaches: \textit{merging} and \textit{growing}. The \textit{merging} protocol involves creating small GHZ states and then ``merging" them into a larger GHZ state via measurements. In contrast, the \textit{growing} protocol is fully unitary and does not involve any measurements. It begins by entangling a small group of neighboring qubits into a GHZ state and then iteratively expands the entanglement to include their neighbors, continuing this process until the desired size is achieved. To compare the two protocols and analyze their respective strengths and limitations, we first implemented both methods on state-of-the-art IBM quantum hardware. Given the limitations of current NISQ devices, we then conducted simulations across various graph topologies inspired by IBM, Google, and distributed quantum computing architectures to gain deeper insights under prospective near-term conditions. We evaluated the performance of the \textit{merging} and \textit{growing} protocols using several figures of merit. Although the \textit{growing} protocol performs better on today’s hardware, our results indicate that the \textit{merging} protocol has the potential to surpass it as two-qubit gate fidelities and readout capabilities continue to improve. We also observe a significant discrepancy between the fidelities of GHZ states obtained from noisy simulations and those measured on actual hardware. The fidelities presented were computed in the absence of error mitigation, and are therefore expected to improve upon the application of appropriate mitigation strategies \cite{koenig2024, Liao2025}.

We have identified several promising directions for future work. While our hardware experiments were conducted exclusively on IBM quantum devices, it would be valuable to extend this study to other architectures such as photonic quantum computers, trapped-ion, and neutral-atom platforms. Since the \textit{merging} protocol can be formulated as a Markov decision process, it is, in principle, possible to apply reinforcement learning to discover more sophisticated strategies for star selection beyond simple heuristics like highest or average degree \cite{rl_simon, rl_sumeet}. Another key area is the development of more accurate methods for estimating the noise models of different hardware systems. Additionally, using specialized SDKs, such as Quantum Rings, could enable more efficient and tailored simulations \cite{quantum_rings}. Further optimization of the protocols under realistic conditions, including node or edge failures and creating entangled states only among preselected nodes also presents an important avenue for research \cite{cka}. We conjecture that many of these layout specific challenges ultimately reduce to optimization problems on graphs.

\section{Acknowledgments}

We thank Peter van Loock, Joachim von Zanthier, Kai Phillip Schmidt, Frank Wilhelm-Mauch, Christian Melzer, and Sumeet Khatri for useful discussions. Additionally, we thank Karl Jansen for valuable support during the completion of this work. We thank Lucas Marti and Pradip Laha for proof reading the manuscript. We received funds by the Deutsche Forschungsgemeinschaft (DFG, German Research Foundation)—Project-ID 429529648—TRR 306 QuCoLiMa (“Quantum Cooperativity of Light and Matter”) and support by its integrated Research Training Group (iRTG). Sumeet thank the support by the Munich Quantum Valley, which is supported by the Bavarian state government with funds from the Hightech Agenda Bayern Plus.

\bibliography{final_references}
\newpage

\onecolumngrid

\appendix

\section{Protocol for merging GHZ states}\label{appendix_heurstic_protocol}
   
    In this section, we present a protocol for merging two GHZ states.
    
    Consider the following two GHZ states:
    \begin{align}
        \ket{\text{GHZ}_{n+1}}_{A_1B_{1:n}}&=\frac{1}{\sqrt{2}}(\ket{0}_{A_1}\otimes\ket{0}_{B_{1:n}}^{\otimes n}+\ket{1}_{A_1}\otimes\ket{1}_{B_{1:n}}^{\otimes n}),\\
        \ket{\text{GHZ}_{m+1}}_{A_2C_{1:m}}&=\frac{1}{\sqrt{2}}(\ket{0}_{A_2}\otimes\ket{0}_{C_{1:m}}^{\otimes m}+\ket{1}_{A_2}\otimes\ket{1}_{C_{1:m}}^{\otimes m}),
    \end{align}
    where $A_1$ and $A_2$ are the centers of the two GHZ states, and we have used the abbreviation $B_{1:n}\equiv B_1B_2\dotsb B_n$ (similarly for $C_{1:m}$). The protocol for merging these two states into a larger GHZ state is as follows:
    \begin{enumerate}
        \item Perform the gate $\text{$CX$}_{A_1A_2}$ between the qubits $A_1$ and $A_2$.
        \item Measure the qubit $A_2$ in the $Z$-basis (i.e., the $\{\ket{0},\ket{1}\}$ basis).
        \item Communicate the outcome $x\in\{0,1\}$ of the measurement to the nodes $C_1,C_2,\dotsc,C_m$.
        \item \begin{enumerate} \item If the outcome is $x=0$, then $C_1,C_2,\dotsc,C_m$ do nothing. \item If the outcome is $x=1$, then $C_1,C_2,\dotsc,C_m$ apply the Pauli-$X$ gate to their qubits. \end{enumerate}
    \end{enumerate}
    
    Let us now prove that after this protocol we deterministically have the GHZ state $\ket{\text{GHZ}_{n+m+1}}_{A_1B_{1:n}C_{1:m}}$ shared by $A_1,B_1,\dotsc,B_n,C_1,\dotsc,C_m$. First,
    \begin{align}
        \ket{\text{GHZ}_{n+1}}\otimes\ket{\text{GHZ}_{m+1}}&=\frac{1}{2}(\ket{0}_{A_1}\otimes\ket{0}_{A_2}\otimes\ket{0}_{B_1^n}^{\otimes n}\otimes\ket{0}_{C_1^m}^{\otimes m} \nonumber\\
        &\qquad + \ket{0}_{A_1}\otimes\ket{1}_{A_2}\otimes\ket{0}_{B_1^n}^{\otimes n}\otimes\ket{1}_{C_1^m}^{\otimes m} \nonumber\\
        &\qquad + \ket{1}_{A_1}\otimes\ket{0}_{A_2}\otimes\ket{1}_{B_1^n}^{\otimes n}\otimes\ket{0}_{C_1^m}^{\otimes m} \nonumber\\
        &\qquad + \ket{1}_{A_1}\otimes\ket{1}_{A_2}\otimes\ket{1}_{B_1^n}^{\otimes n}\otimes\ket{1}_{C_1^m}^{\otimes m}).
    \end{align}
    Then, after the $CX$ between $A_1$ and $A_2$, the state becomes
    \begin{align}
        &\frac{1}{2}(\ket{0}_{A_1}\otimes\ket{0}_{A_2}\otimes\ket{0}_{B_1^n}^{\otimes n}\otimes\ket{0}_{C_1^m}^{\otimes m} \nonumber\\
        &\quad + \ket{0}_{A_1}\otimes\ket{1}_{A_2}\otimes\ket{0}_{B_1^n}^{\otimes n}\otimes\ket{1}_{C_1^m}^{\otimes m} \nonumber\\
        &\quad + \ket{1}_{A_1}\otimes\ket{1}_{A_2}\otimes\ket{1}_{B_1^n}^{\otimes n}\otimes\ket{0}_{C_1^m}^{\otimes m} \nonumber\\
        &\quad + \ket{1}_{A_1}\otimes\ket{0}_{A_2}\otimes\ket{1}_{B_1^n}^{\otimes n}\otimes\ket{1}_{C_1^m}^{\otimes m}).
    \end{align}
    From this, we can see immediately that, when $A_2$ measures its qubit in the $\{\ket{0},\ket{1}\}$ basis, then the outcome ``0'' occurs with probability $\frac{1}{2}$, and the state conditioned on this outcome is
    \begin{equation}
        \frac{1}{\sqrt{2}}(\ket{0}_{A_1}\otimes\ket{0}_{B_1^n}^{\otimes n}\otimes\ket{0}_{C_1^m}^{\otimes m}+\ket{1}_{A_1}\otimes\ket{1}_{B_1^n}^{\otimes n}\otimes\ket{1}_{C_1^m}^{\otimes m})=\ket{\text{GHZ}_{n+m+1}}.
    \end{equation}
    On the other hand, if the outcome ``1'' occurs, also with probability $\frac{1}{2}$, then the state conditioned on this outcome is
    \begin{equation}
        \frac{1}{\sqrt{2}}(\ket{0}_{A_1}\otimes\ket{0}_{B_1^n}^{\otimes n}\otimes\ket{1}_{C_1^m}^{\otimes m}+\ket{1}_{A_1}\otimes\ket{1}_{B_1^n}^{\otimes n}\otimes\ket{0}_{C_1^m}^{\otimes m}).
    \end{equation}
    Thus, if $C_1,C_2,\dotsc,C_m$ apply the Pauli $X$ gate to their qubits, then we obtain the required GHZ state $\ket{\text{GHZ}_{n+m+1}}$. This completes the proof.

\section{Evaluating different star selection criteria in the \textit{merging} protocol}
\label{app:eval_star_selec_crit}

As already noted in Section~\ref{sec:merging}, it is in principle possible to pick stars, in step 1 of the \textit{merging} protocol, based on other criteria than the highest node degree. In this section, we investigated this possibility by picking stars based on target node degrees of the star center above, below and equal to the average degree of the input graph. Additionally, we also pick stars for the IBM and rectangular layouts based on absolute values of the target star size. Setting a target node degree is handled by the scaling factor, which relates the target degree to the average degree (cf. Section~\ref{sec:merging}). In the following, we will evaluate the different criteria of picking stars for the considered layout topologies (IBM, rectangular and random), to find the best criteria for each layout topology.\\

\begin{figure*}[htb!]
\begin{tabular}{ccc}
    \includegraphics[width=0.3\textwidth]{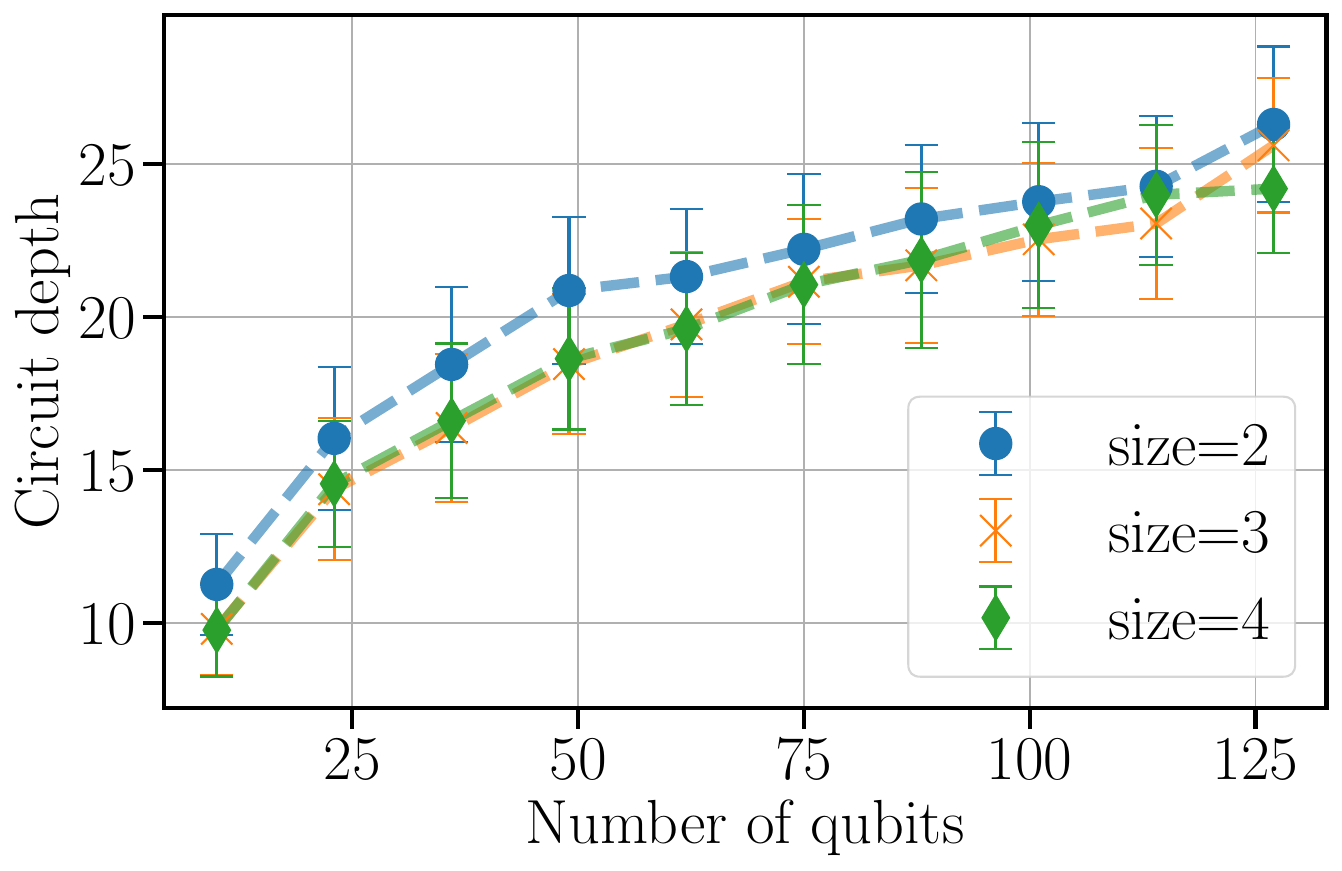}&
    \includegraphics[width=0.3\textwidth]{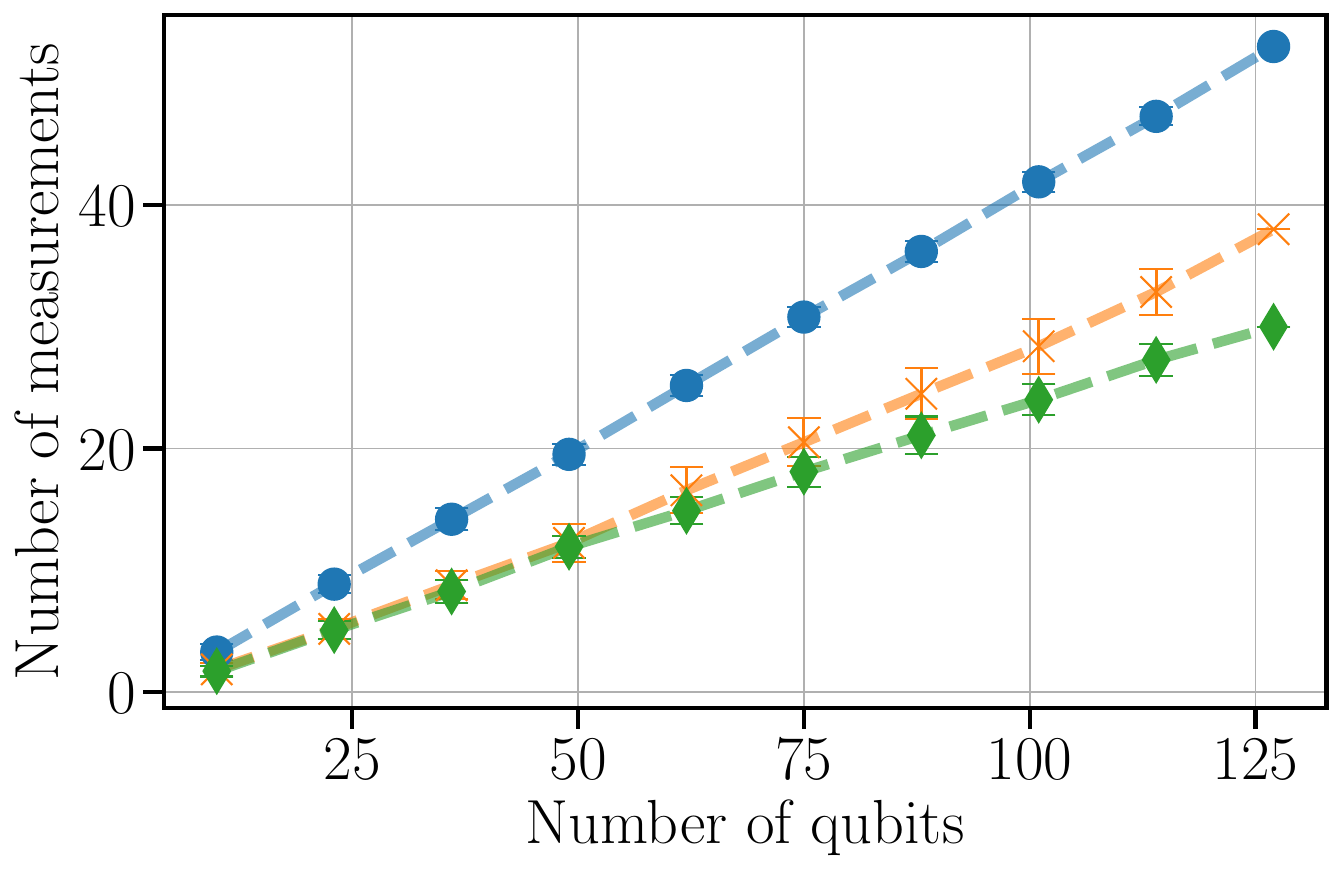}&
    \includegraphics[width=0.3\textwidth]{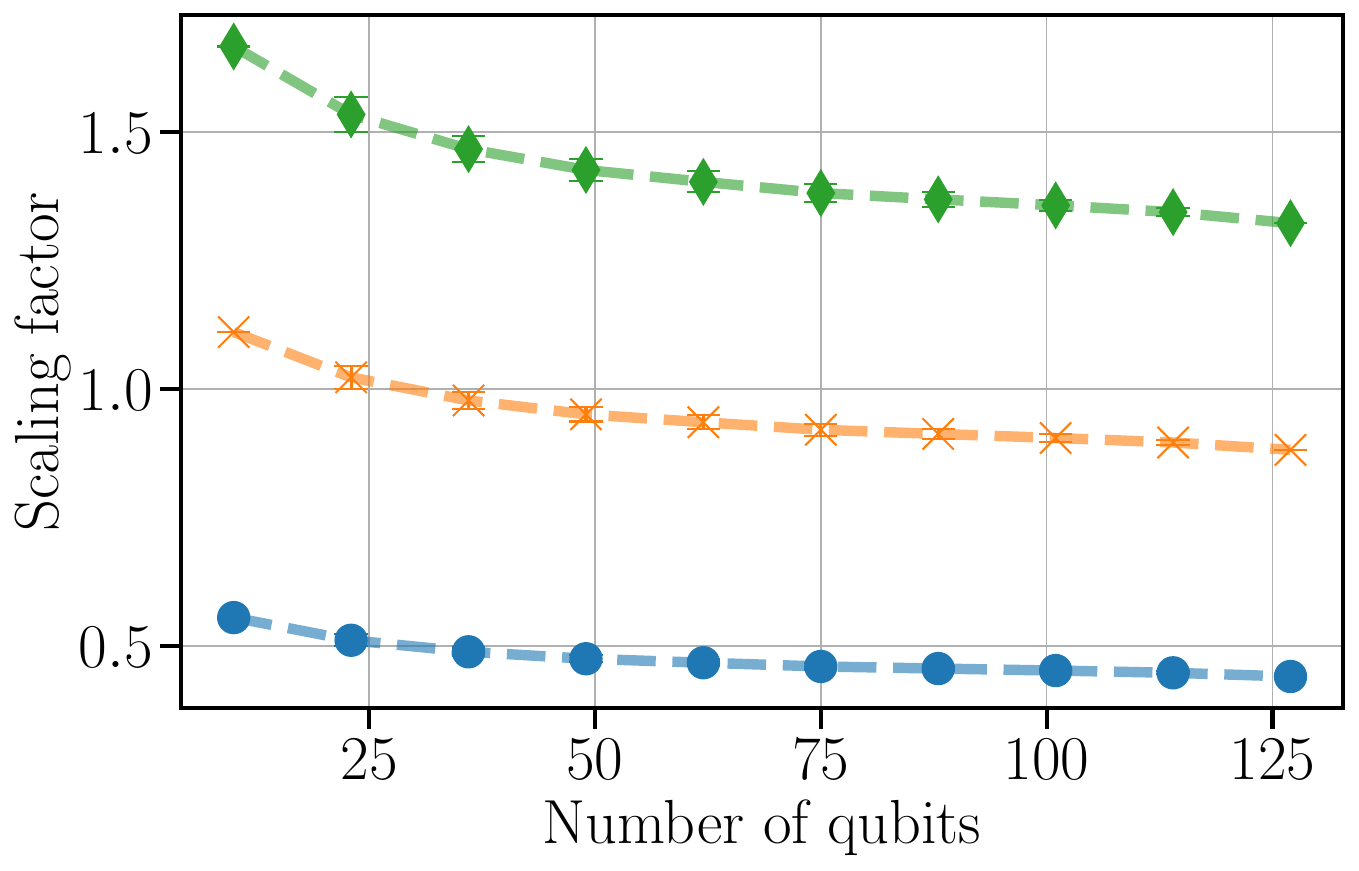}\\
    (a) & (b) & (c)\\
    \end{tabular}
    \caption{Random subgraph sampling from IBM's 127-qubit Eagle layout \cite{IBM_Eagle}. Here we compare our \textit{merging} protocol with varying size for the star selection. The curves represent a star selection based on absolute target sizes of $2$ (blue circles), $3$ (orange crosses) and $4$ nodes (green diamonds). Panels (a), (b) and (c) show the averaged circuit depth, number of measurements and scaling factor in dependence of the number of qubits in the final GHZ state. For each GHZ state size, we generated $100$ subgraph samples of this size randomly from the initial layout graph. The error bars show the standard deviation obtained from averaging over the samples.}
    \label{fig:ibm_brisbane_substate_size}
\end{figure*}

\begin{figure*}[htb!]
\begin{tabular}{ccc}
    \includegraphics[width=0.3\textwidth]{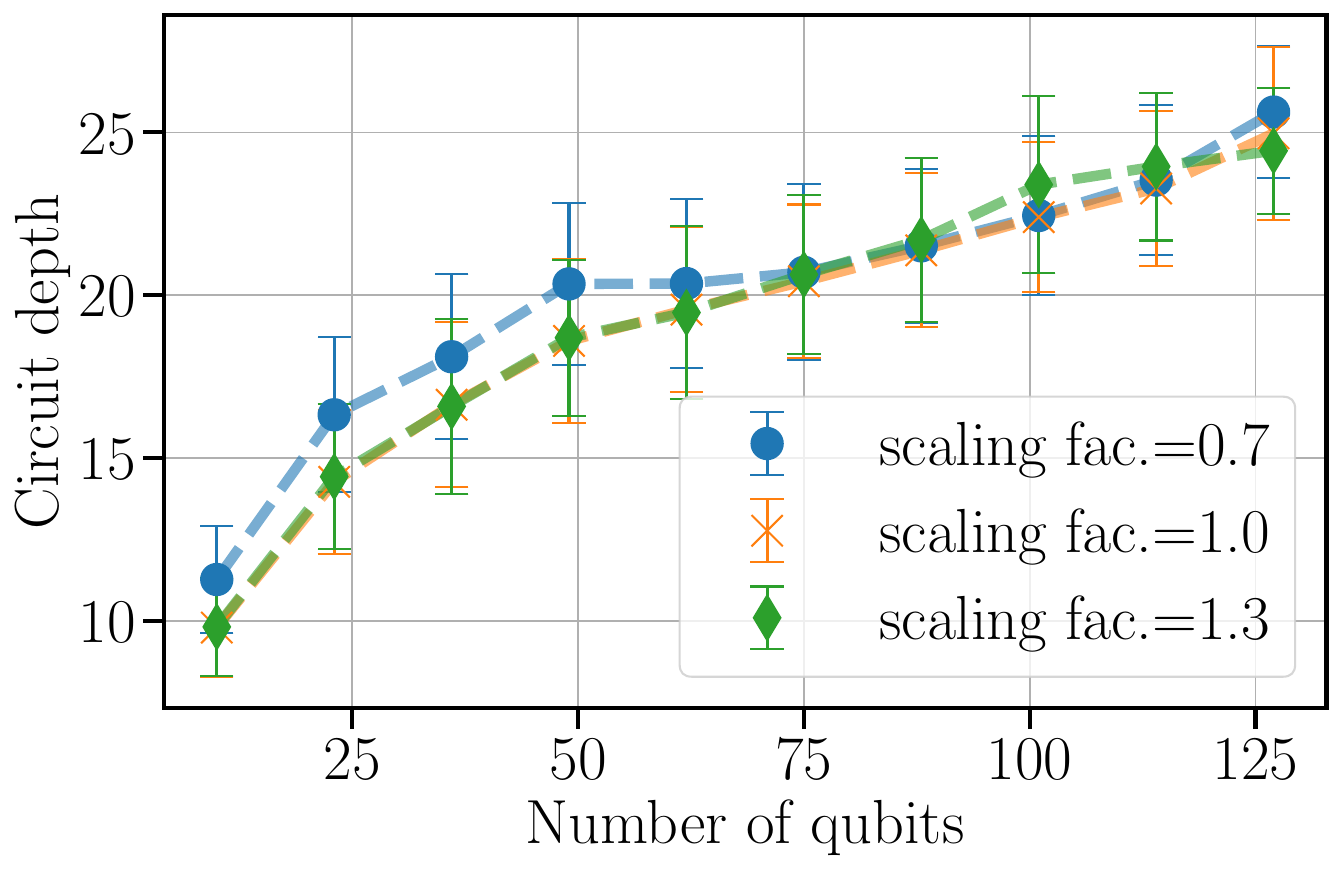}&
    \includegraphics[width=0.3\textwidth]{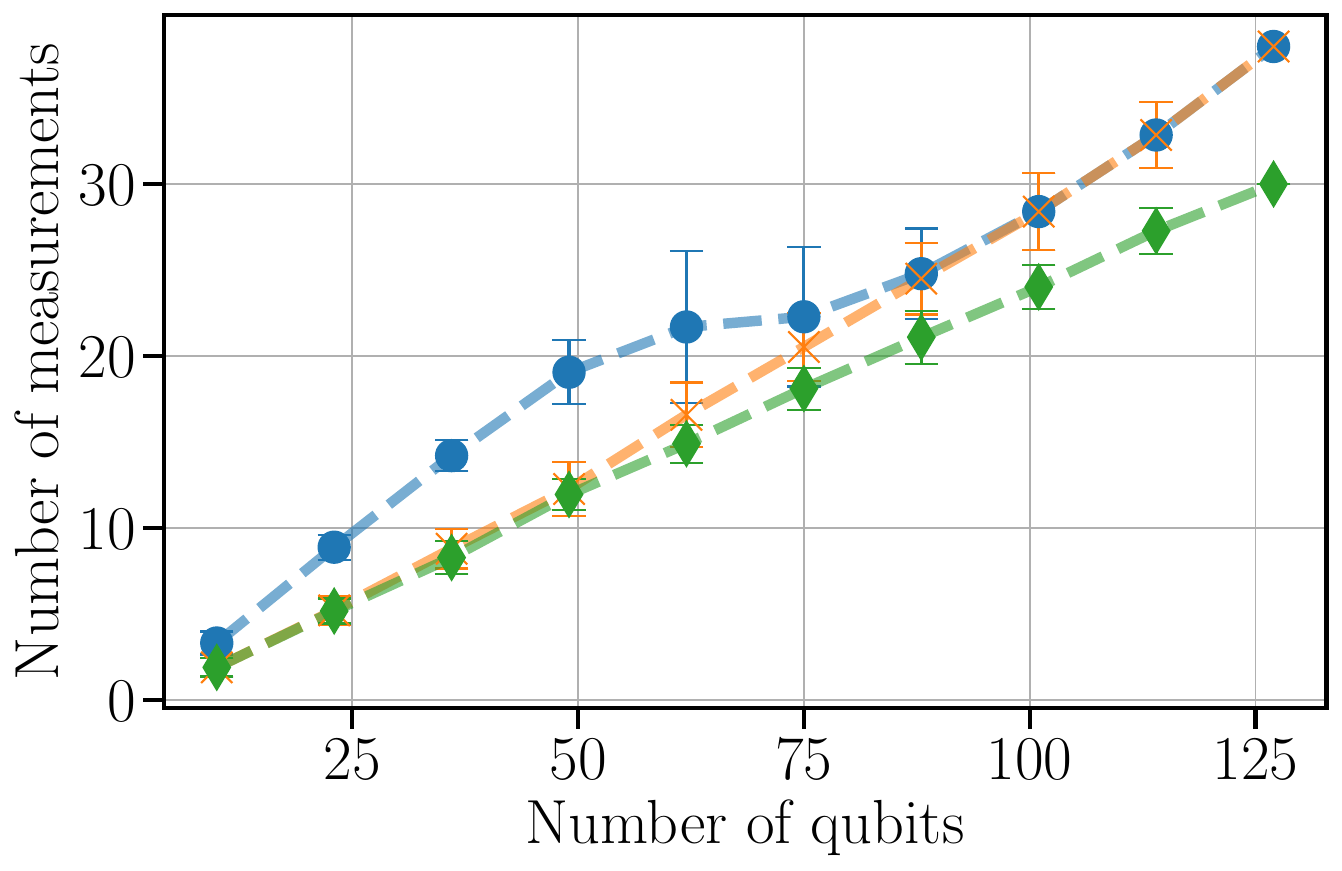}&
    \includegraphics[width=0.3\textwidth]{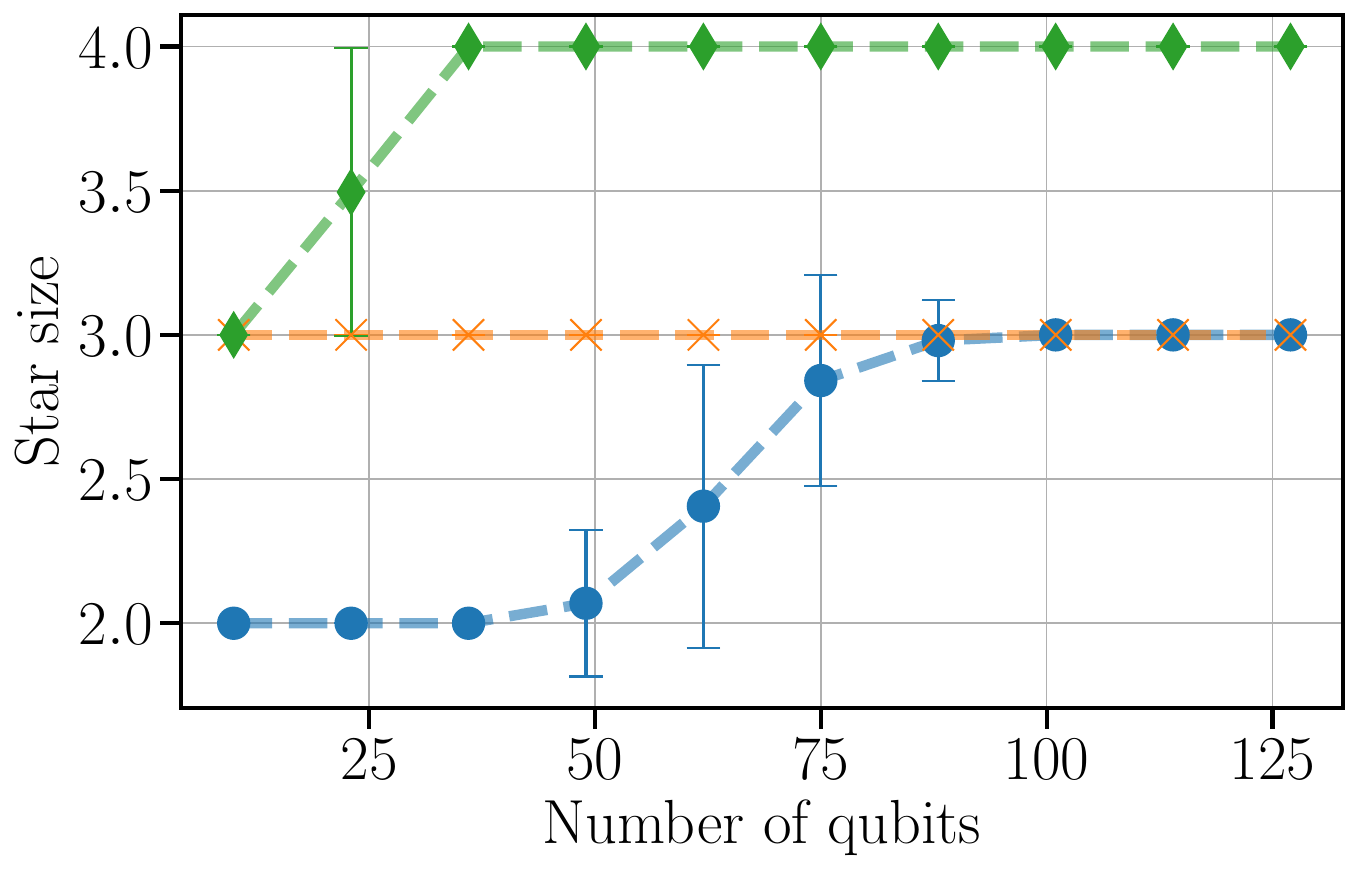}\\
    (a) & (b) & (c) \\
    \end{tabular}
    \caption{Random subgraph sampling from IBM's 127-qubit Eagle layout \cite{IBM_Eagle}. Here we compare our \textit{merging} protocol with varying size for the star selection. The curves represent a star selection based on scaling factors of $0.7$ (blue circles), $1.0$ (orange crosses) and $1.3$ (green diamonds). Panels (a), (b) and (c) show the averaged circuit depth, number of measurements and star size in dependence of the number of qubits in the final GHZ state. For each GHZ state size, we generated $100$ subgraph samples of this size randomly from the initial layout graph. The error bars show the standard deviation obtained from averaging over the samples.}
    \label{fig:ibm_brisbane_substate_size_fac}
\end{figure*}

\begin{figure*}[htb!]
\begin{tabular}{ccc}
    \includegraphics[width=0.3\textwidth]{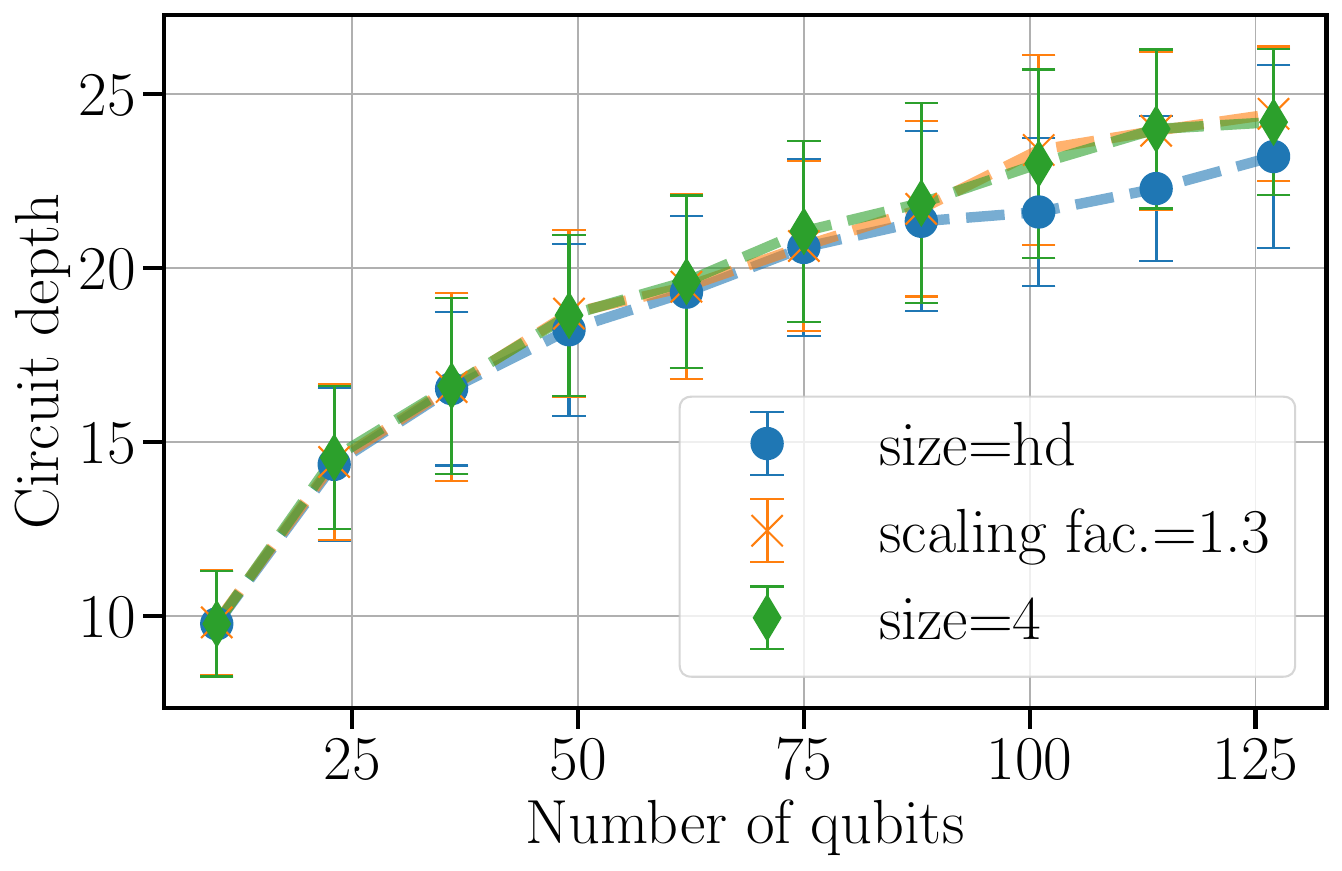}&
    \includegraphics[width=0.3\textwidth]{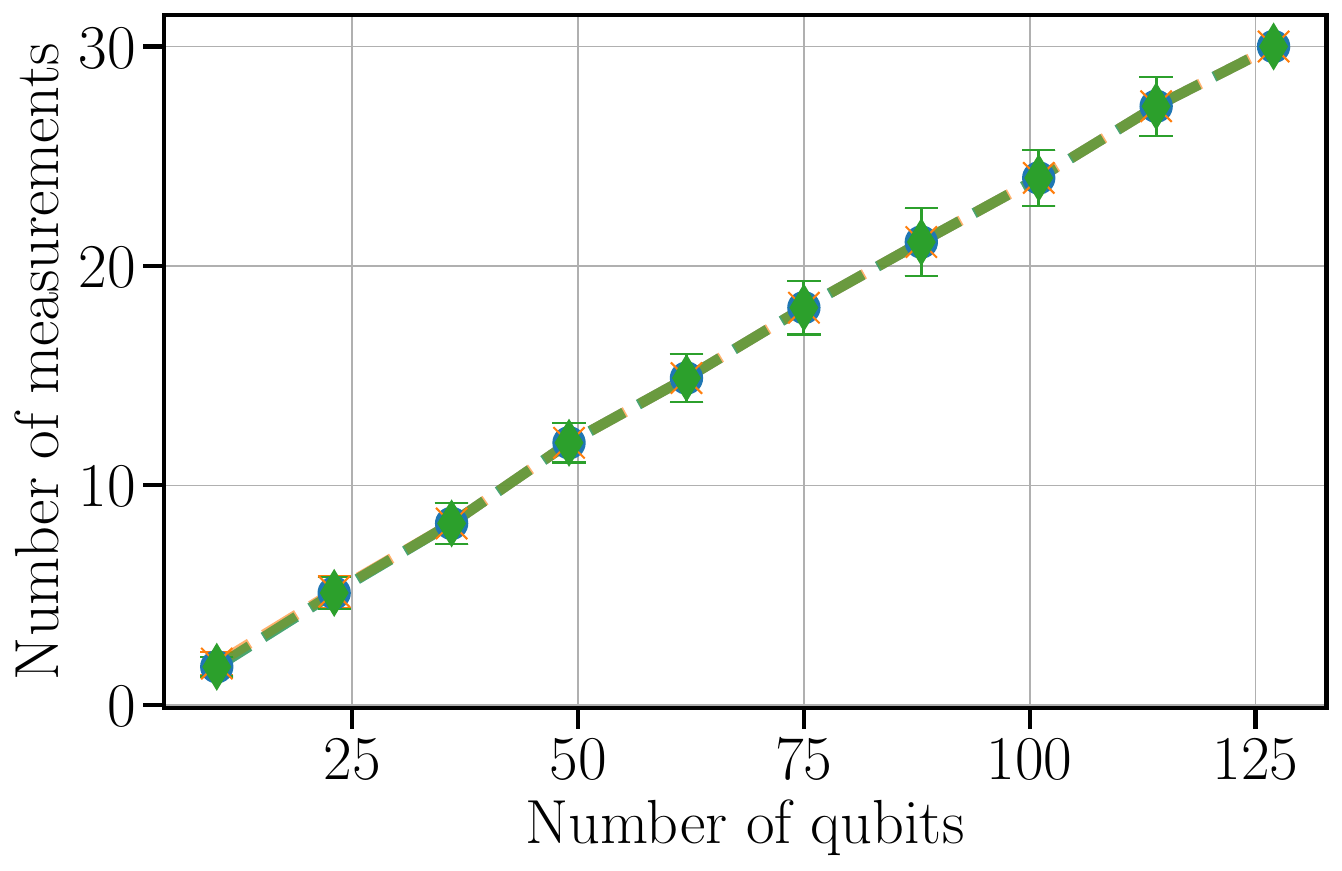}&
    \includegraphics[width=0.3\textwidth]{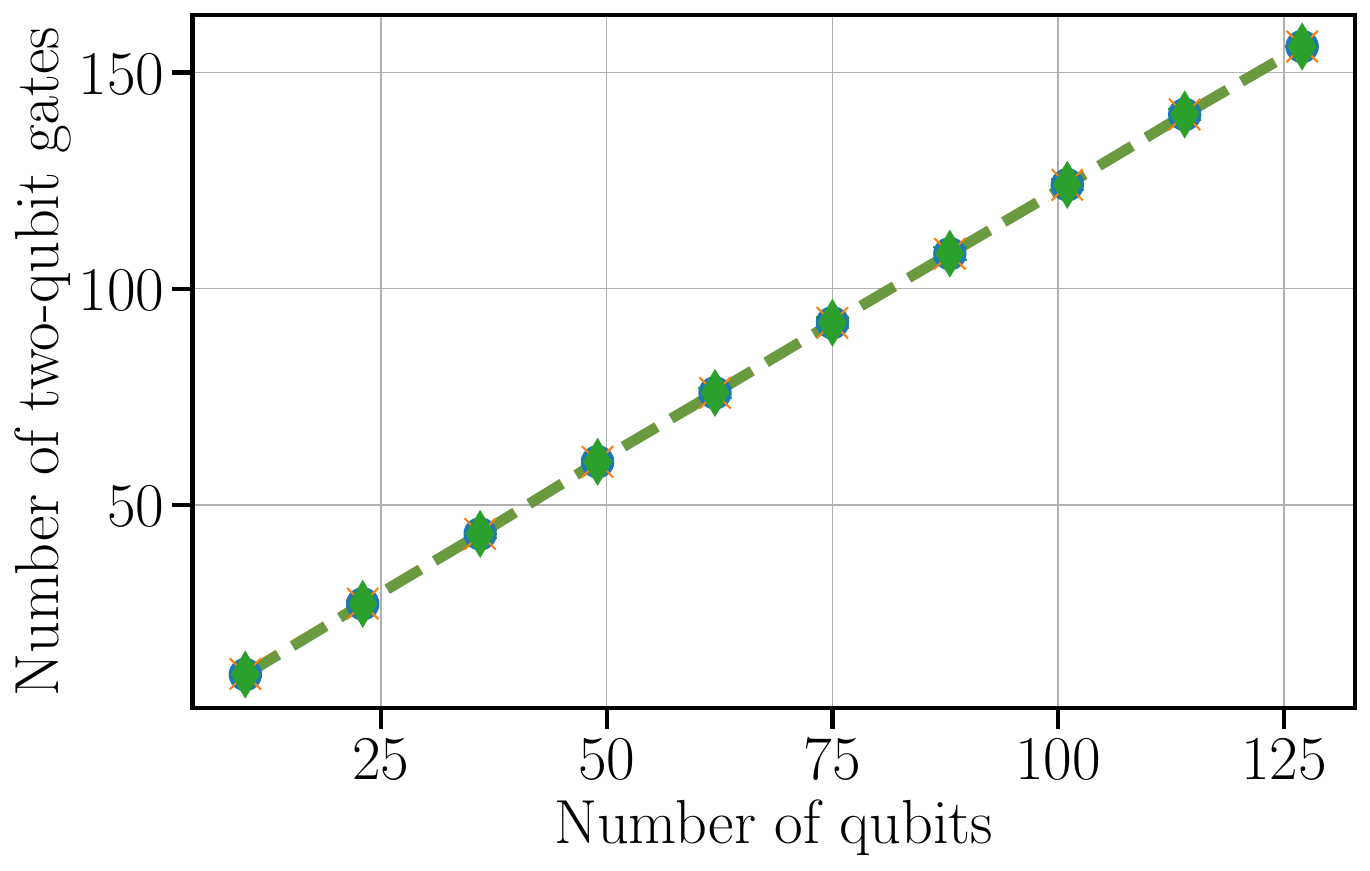}\\
       (a) & (b) & (c) \\
    \end{tabular}
    \caption{Random subgraph sampling from IBM's 127-qubit Eagle layout \cite{IBM_Eagle}. Here we compare our \textit{merging} protocol with varying size for the star selection. The curves represent a star selection based on the highest degree (blue circles), a scaling factor of $1.3$ (orange crosses) and an absolute target size of $4$ nodes (green diamonds). Panels (a), (b) and (c) show the averaged circuit depth, number of measurements and number of two-qubit gates in dependence of the number of qubits in the final GHZ state. For each GHZ state size, we generated $100$ subgraph samples of this size randomly from the initial layout graph.  The error bars show the standard deviation obtained from averaging over the samples.}
    \label{fig:ibm_brisbane_substate_size_vs_fac}
    
\end{figure*}

For the considered IBM 127-qubit Eagle layout, the best ways to pick stars in the \textit{merging} protocol is based on the highest degree. This criteria results in the smallest number of measurements, as the number of merges is smallest in that case. For the averaged circuit depth, we cannot determine a significant difference between the tested criteria, although we observe a tendency towards smaller circuit depth for larger star sizes. In Figure~\ref{fig:ibm_brisbane_substate_size}, we varied the absolute size of the selected stars and find that the largest size has the smallest number of measurements. A similar observation can be made for picking the stars based on a target node degree (cf. Figure~\ref{fig:ibm_brisbane_substate_size_fac}). We find that the highest scaling factor results in the smallest number of measurements. From Figure~\ref{fig:ibm_brisbane_substate_size} and Figure~\ref{fig:ibm_brisbane_substate_size_fac} we can conclude that, the larger the size of the selected stars is, the less measurements we have to do while still producing the smallest circuit depth. This shows that the circuit depth reduction we expected from a star generation with equal star sizes is insignificant compared to the increase in the circuit depth due to the higher number of merges. In order to always pick the largest stars, we have to pick stars with the highest degree. We compared the highest tested absolute size and the highest scaling factor to the highest degree criteria in Figure~\ref{fig:ibm_brisbane_substate_size_vs_fac}. Here we don't see any significant differences in the number of measurements, number of two-qubit gates or circuit depth. This can be explained by looking at the IBM layout graph (cf. Figure~\ref{fig:ibm_layout}). The highest degree that occurs in the graph is three, corresponding to a target star size of four. This is already the target size for the largest tested size in Figure~\ref{fig:ibm_brisbane_substate_size} and for the largest scaling factor in Figure~\ref{fig:ibm_brisbane_substate_size_fac}. Hence, there is no difference anymore in the star selection compared to directly picking the nodes with the highest degree.

\begin{figure*}[htb!]
\begin{tabular}{ccc}
    \includegraphics[width=0.3\textwidth]{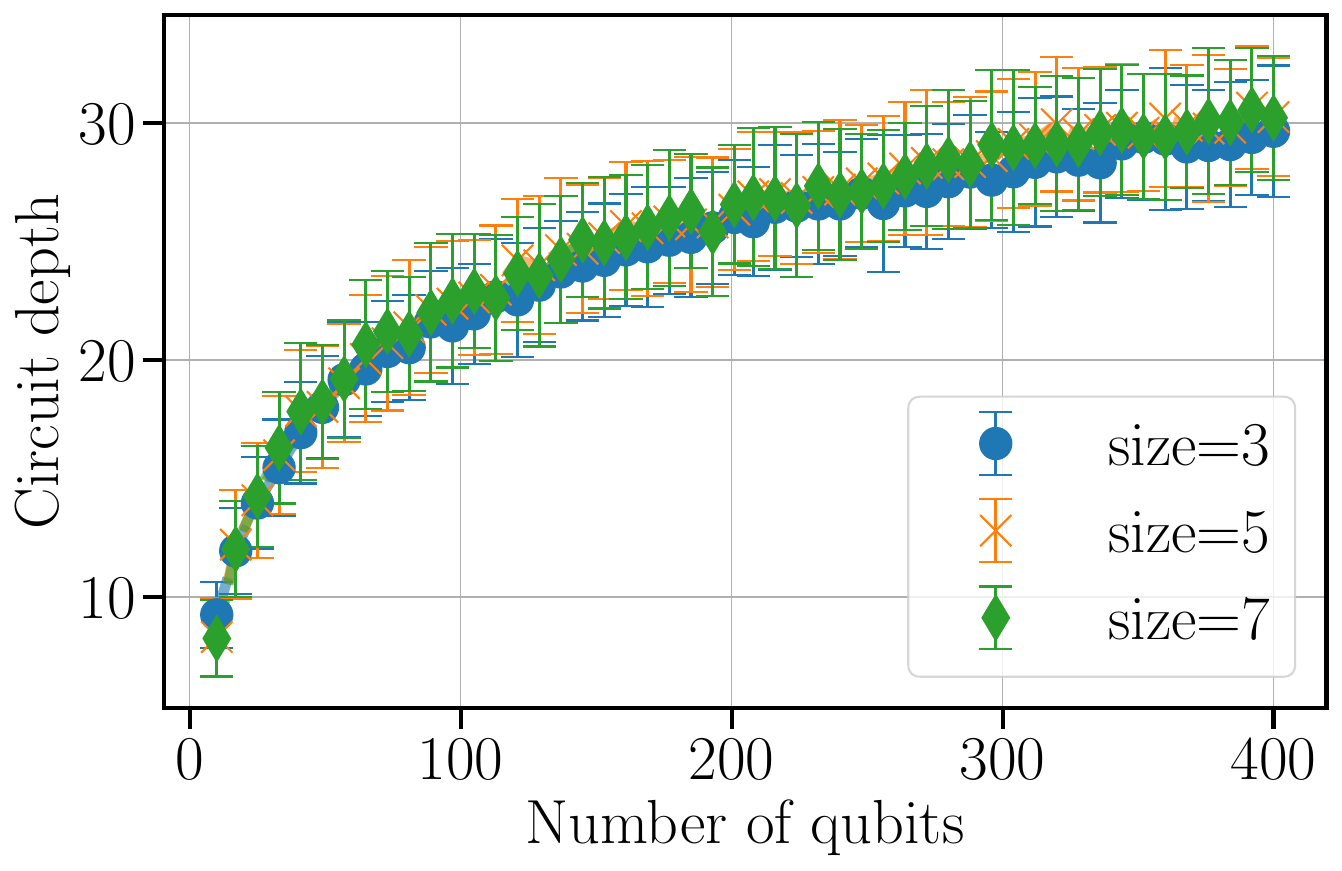}&
    \includegraphics[width=0.3\textwidth]{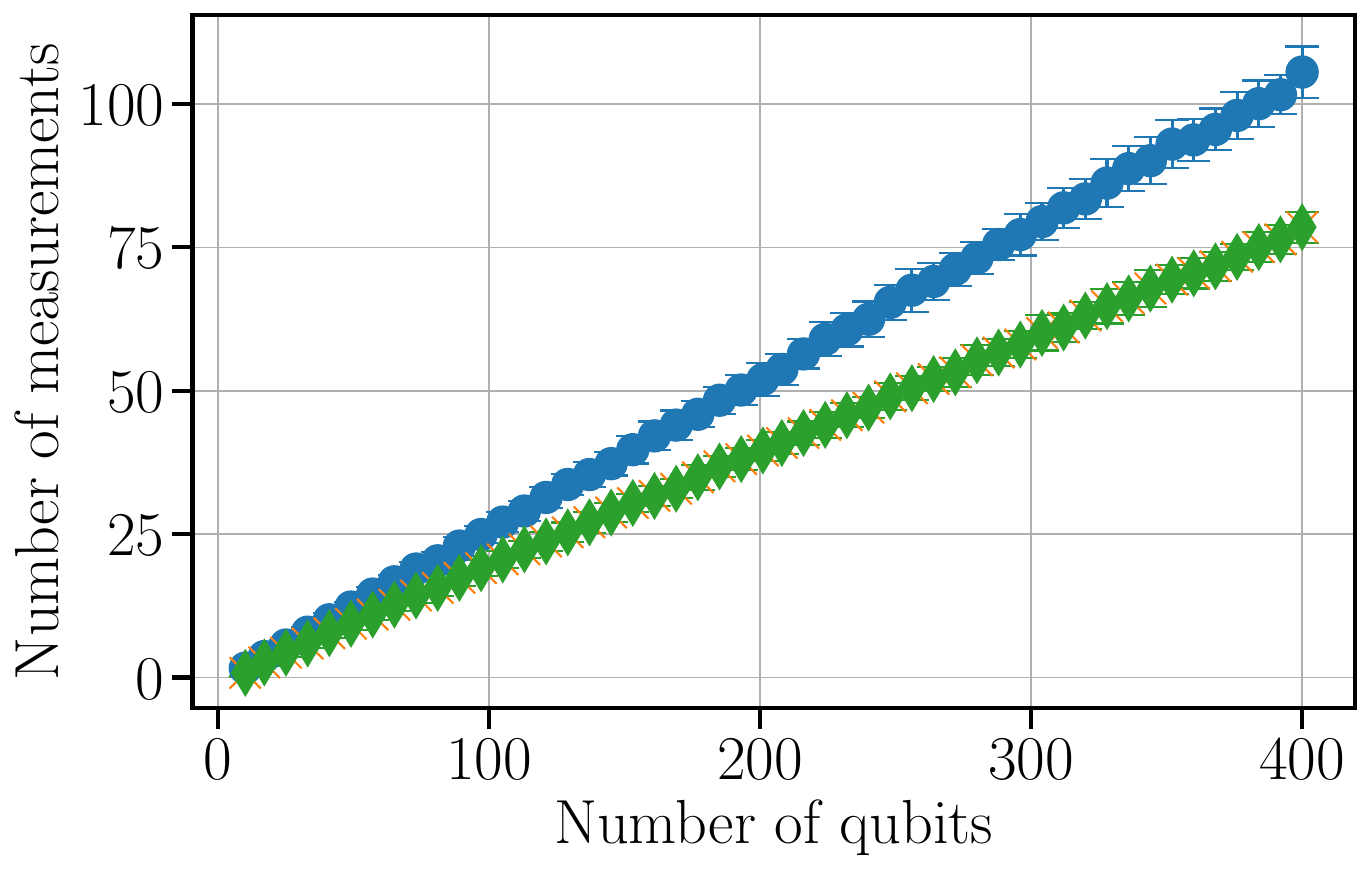}&
    \includegraphics[width=0.3\textwidth]{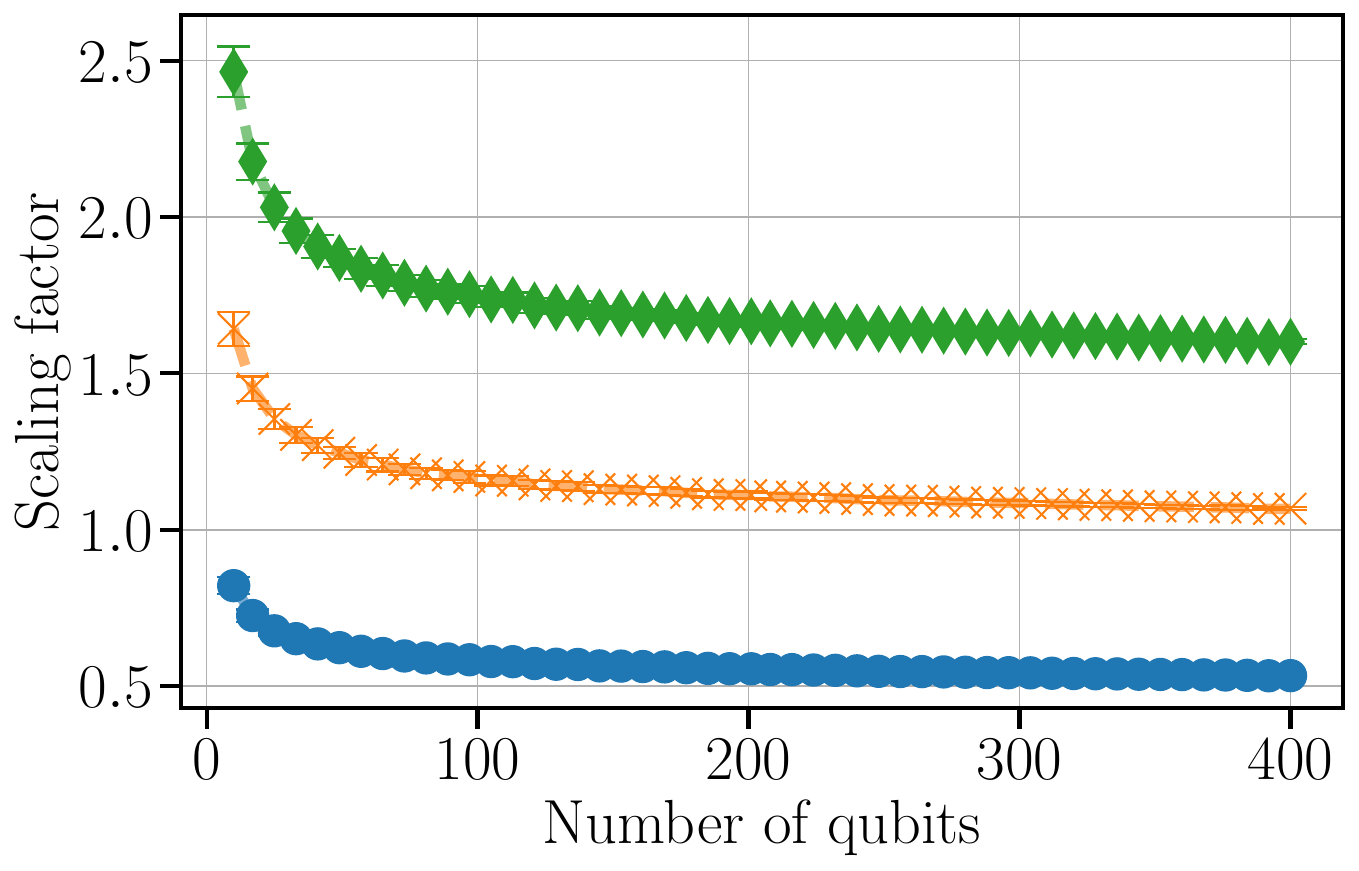}\\

    (a) & (b) & (c) \\
    
\end{tabular}
    \caption{Random subgraph sampling from rectangular grid layout inspired by Google's Willow chip \cite{Willow}. Here we compare our \textit{merging} protocol with varying size for the star selection. The curves represent a star selection based on absolute target sizes of $3$ (blue circles), $5$ (orange crosses) and $7$ nodes (green diamonds). Panels (a), (b) and (c) show the averaged circuit depth, number of measurements and scaling factor in dependence of the number of qubits in the final GHZ state. For each GHZ state size, we generated $100$ subgraph samples of this size randomly from the initial layout graph. The error bars show the standard deviation obtained from averaging over the samples.}
    \label{fig:rect_grid_substate_size}
\end{figure*}

\begin{figure*}[htb!]
\begin{tabular}{ccc}
    \includegraphics[width=0.3\textwidth]{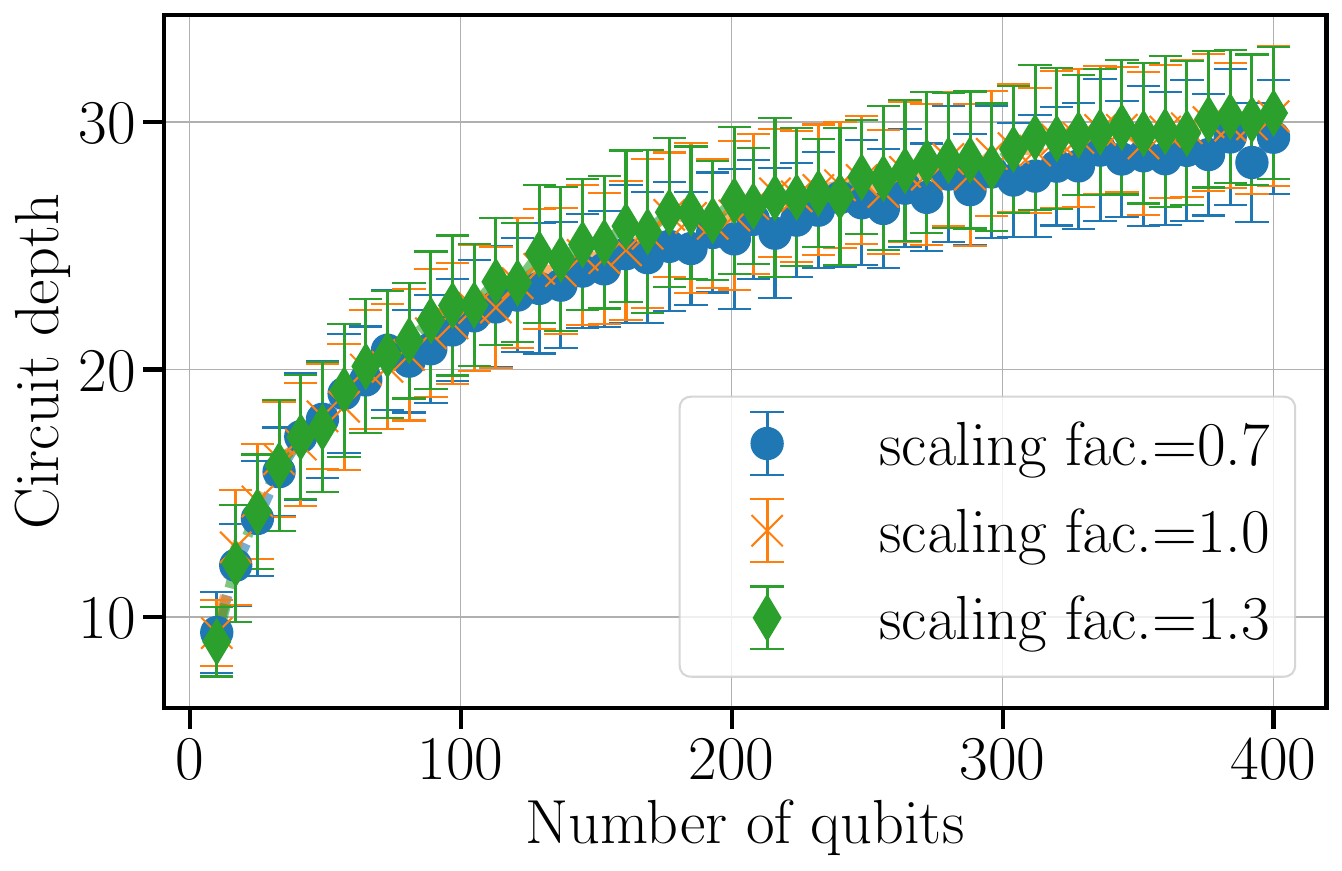}&
    \includegraphics[width=0.3\textwidth]{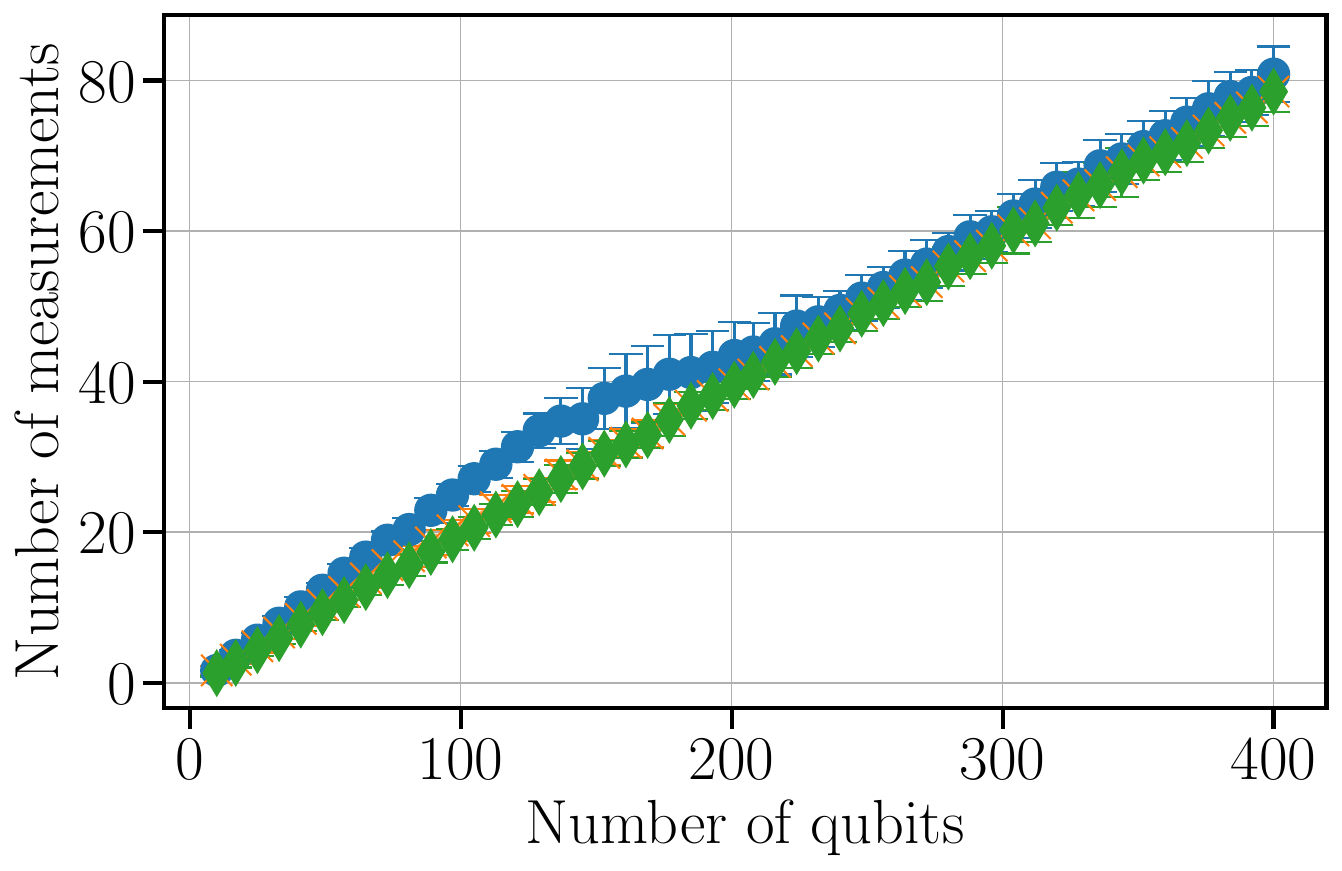}&
    \includegraphics[width=0.3\textwidth]{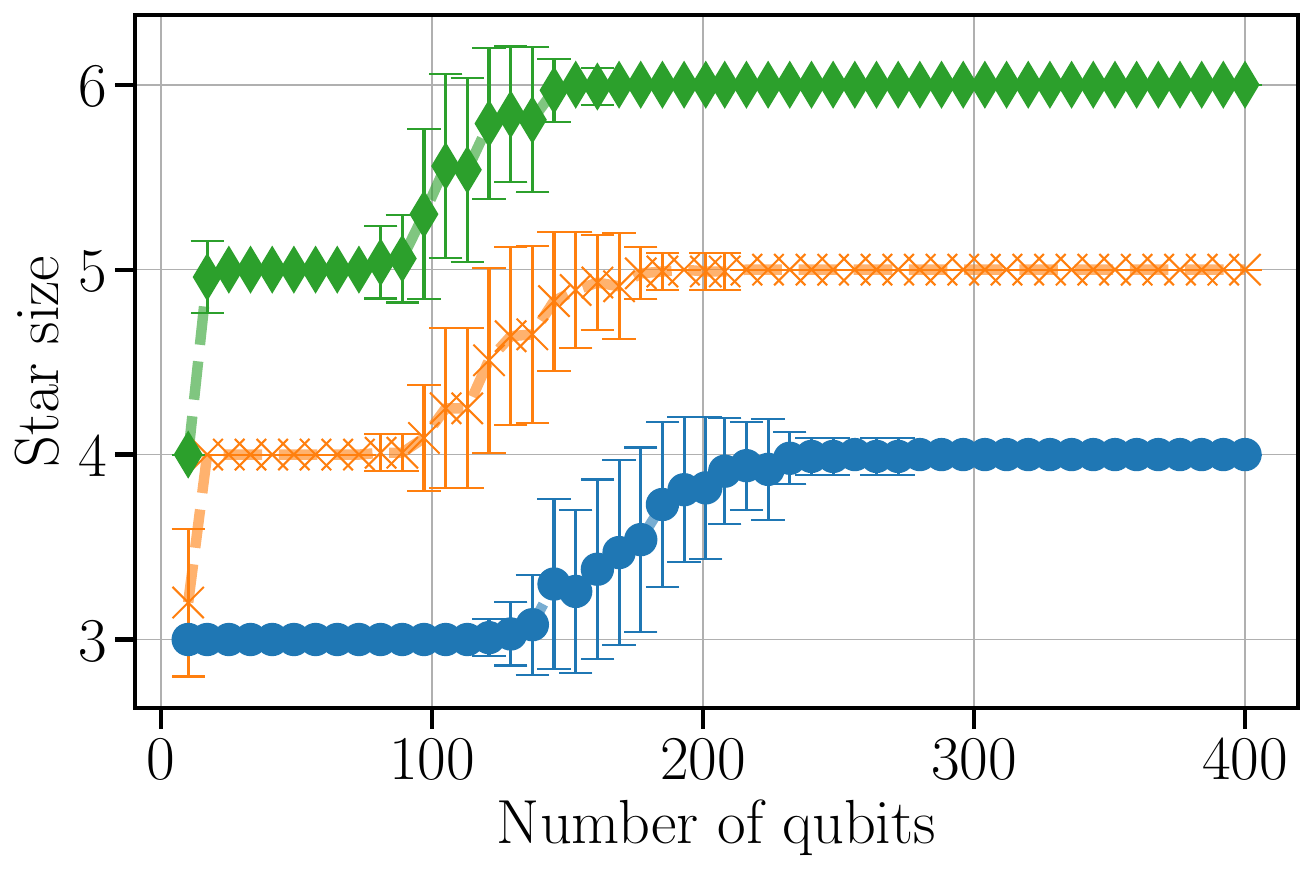}\\
    (a) & (b) & (c) \\
    \end{tabular}
    \caption{Random subgraph sampling from rectangular grid layout inspired by Google's Willow chip \cite{Willow}. Here we compare our \textit{merging} protocol with varying size for the star selection. The curves represent a star selection based on scaling factors of $0.7$ (blue circles), $1.0$ (orange crosses) and $1.3$ (green diamonds). Panels (a), (b) and (c) show the averaged circuit depth, number of measurements and star size in dependence of the number of qubits in the final GHZ state. For each GHZ state size, we generated $100$ subgraph samples of this size randomly from the initial layout graph. The error bars show the standard deviation obtained from averaging over the samples.}
    \label{fig:rect_grid_substate_size_fac}
\end{figure*}

\begin{figure*}[htb!]
\begin{tabular}{ccc}

    \includegraphics[width=0.3\textwidth]{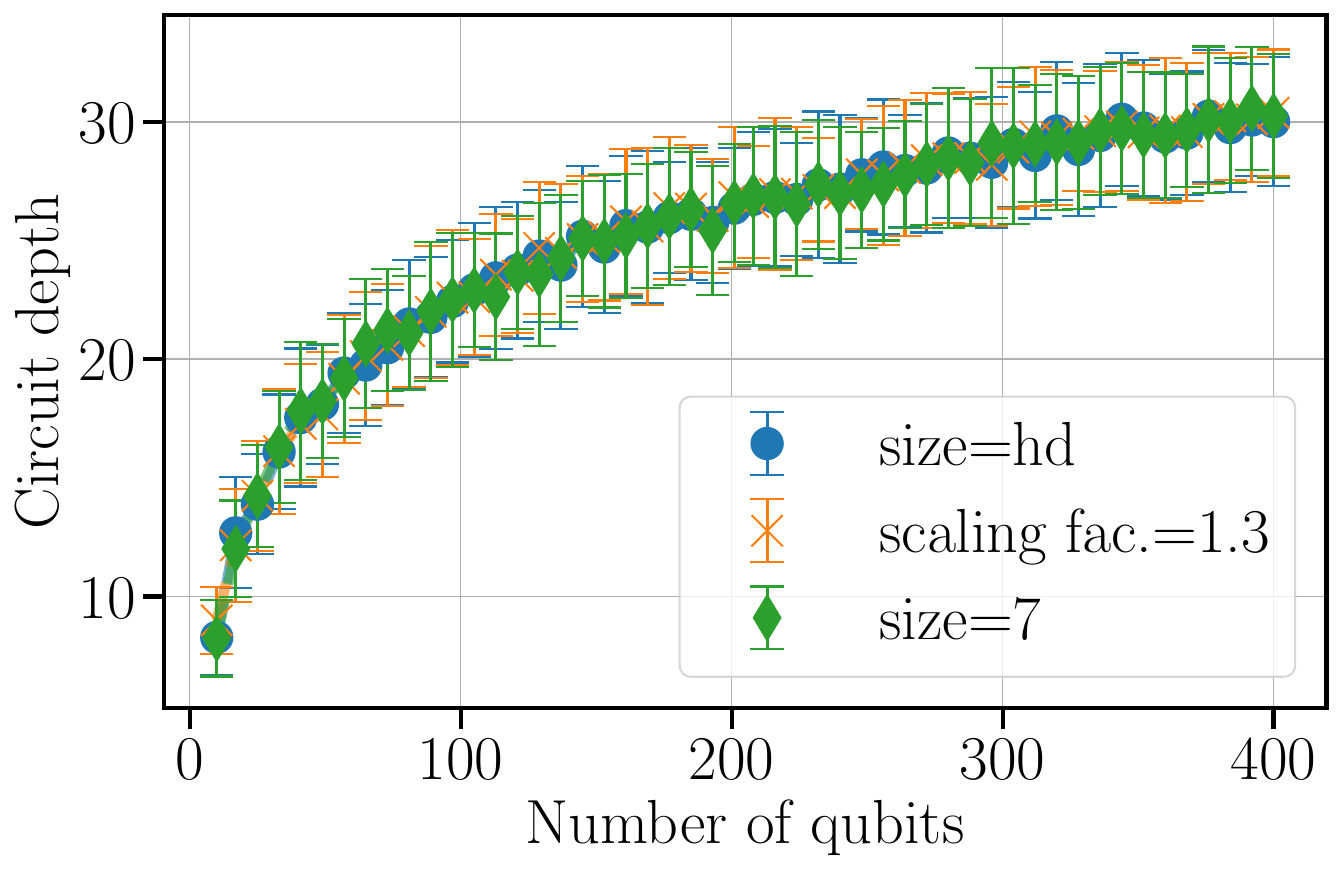}&
    \includegraphics[width=0.3\textwidth]{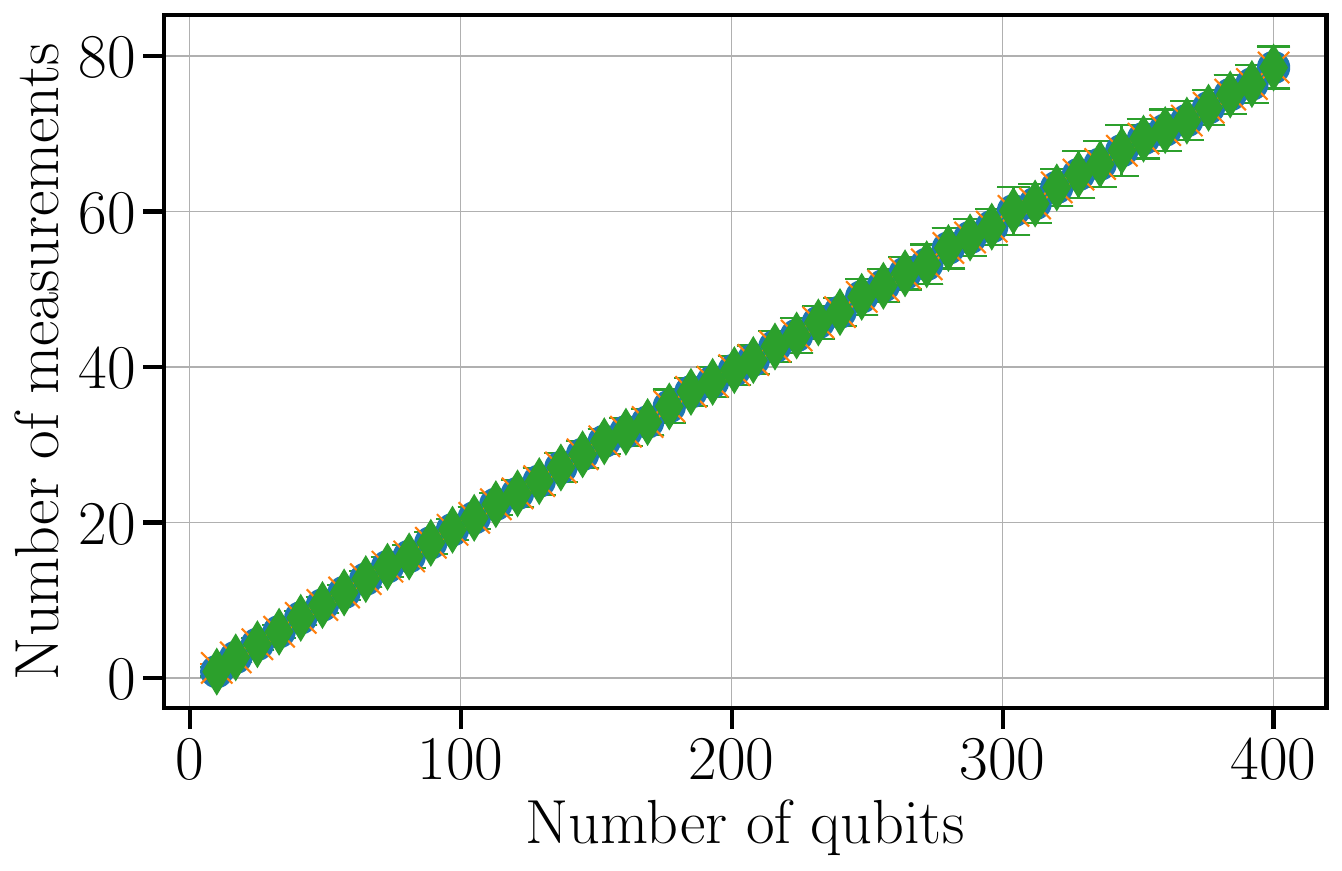}&
    \includegraphics[width=0.3\textwidth]{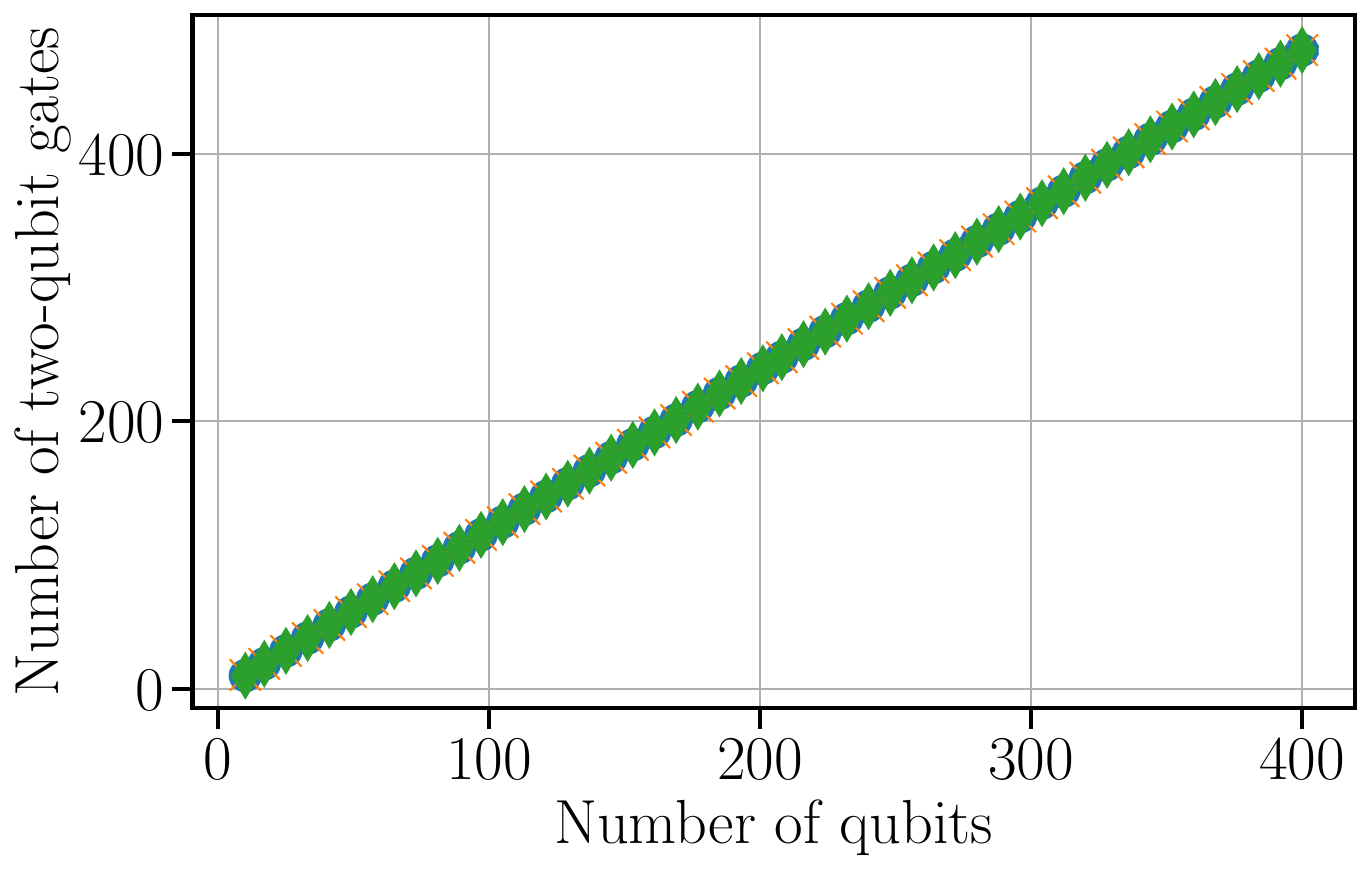}\\
       (a) & (b) & (c) \\
    \end{tabular}

    \caption{Random subgraph sampling from rectangular grid layout inspired by Google's Willow chip \cite{Willow}. Here we compare our \textit{merging} protocol with varying size for the star selection. The curves represent a star selection based on the highest degree (blue circles), a scaling factor of $1.3$ (orange crosses) and an absolute target size of $7$ nodes (green diamonds). Panels (a), (b) and (c) show the averaged circuit depth, number of measurements and number of two-qubit gates in dependence of the number of qubits in the final GHZ state. For each GHZ state size, we generated $100$ subgraph samples of this size randomly from the initial layout graph. The error bars show the standard deviation obtained from averaging over the samples.}
    \label{fig:rect_grid_substate_size_vs_fac}
\end{figure*}

For the rectangular layout, we again find that selecting stars based on the highest degree is the best way, yielding the smallest number of measurements without any significant drawbacks in the averaged circuit depth. Thus, we identify the increasing number of merges as the most dominant factor, if we select stars based on similar star sizes, without any improvements in the circuit depth. We see no difference between the largest absolute star size, the largest scaling factor and the highest degree. Additionally, in this case there is also no difference to the average degree and the second largest absolute size (cf. Figures~\ref{fig:rect_grid_substate_size} and~\ref{fig:rect_grid_substate_size_fac}). Similar to the IBM layout, the reason for this behavior is that the highest node degree in a rectangular grid graph is four which corresponds to a target star size of five. The two largest absolute values in Figure~\ref{fig:rect_grid_substate_size} as well as all target node degrees equal or above the average degree in Figure~\ref{fig:rect_grid_substate_size_fac} already result in a target size of five for the selected stars. Thus, we see no differences in the curves of Figure~\ref{fig:rect_grid_substate_size_vs_fac}.

\begin{figure*}
\begin{tabular}{ccc}

    \includegraphics[width=0.3\textwidth]{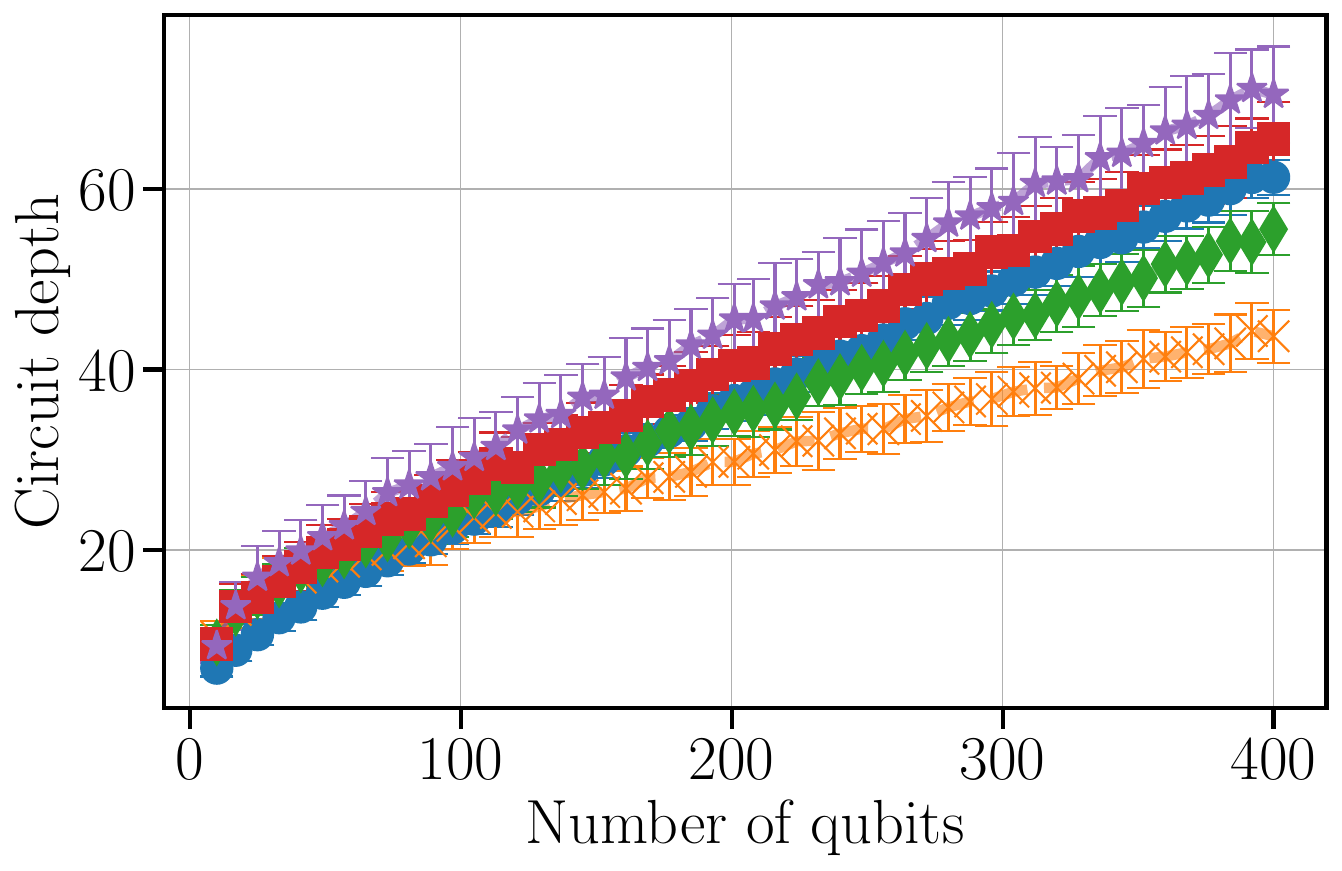}&
    \includegraphics[width=0.3\textwidth]{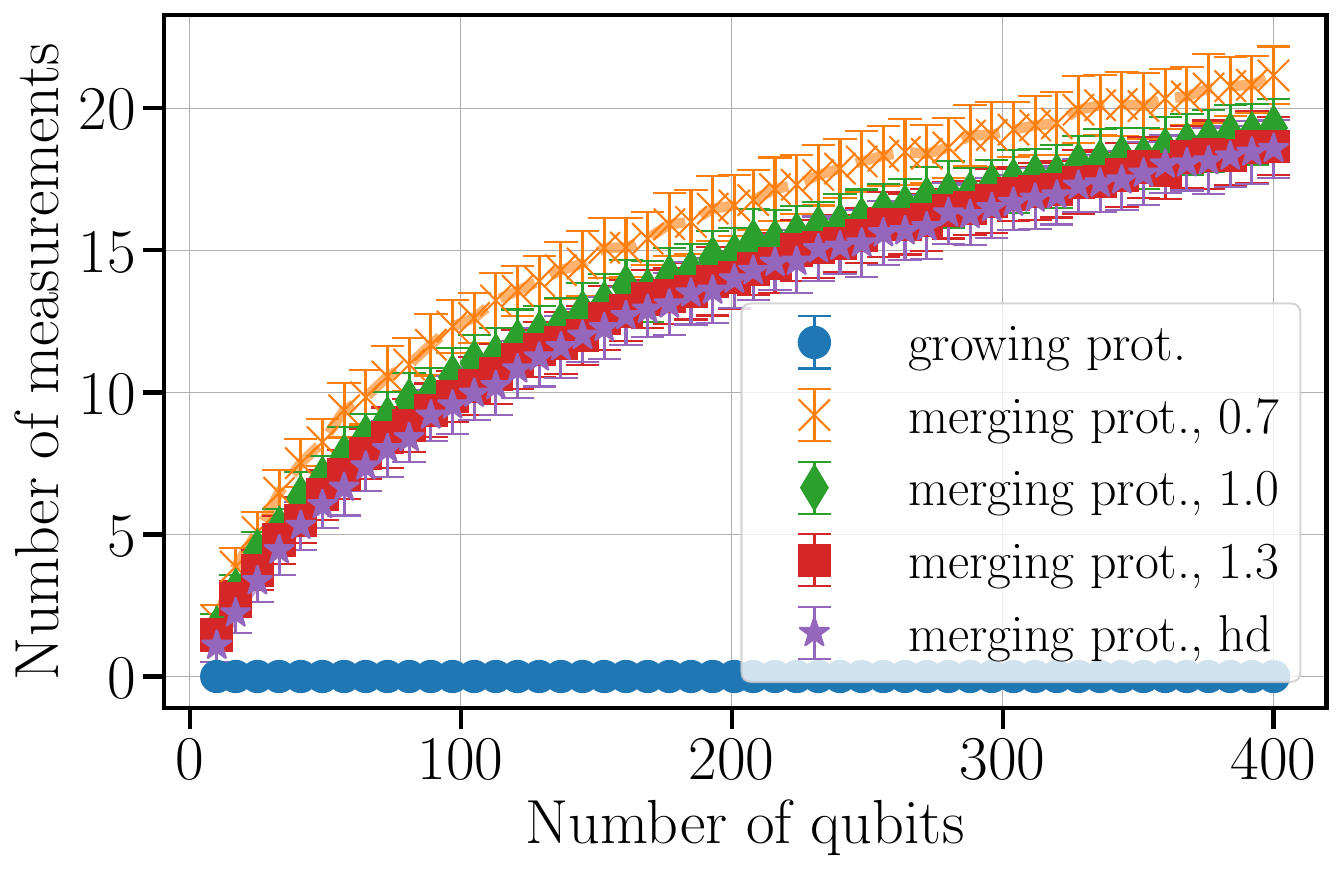}&
    \includegraphics[width=0.3\textwidth]{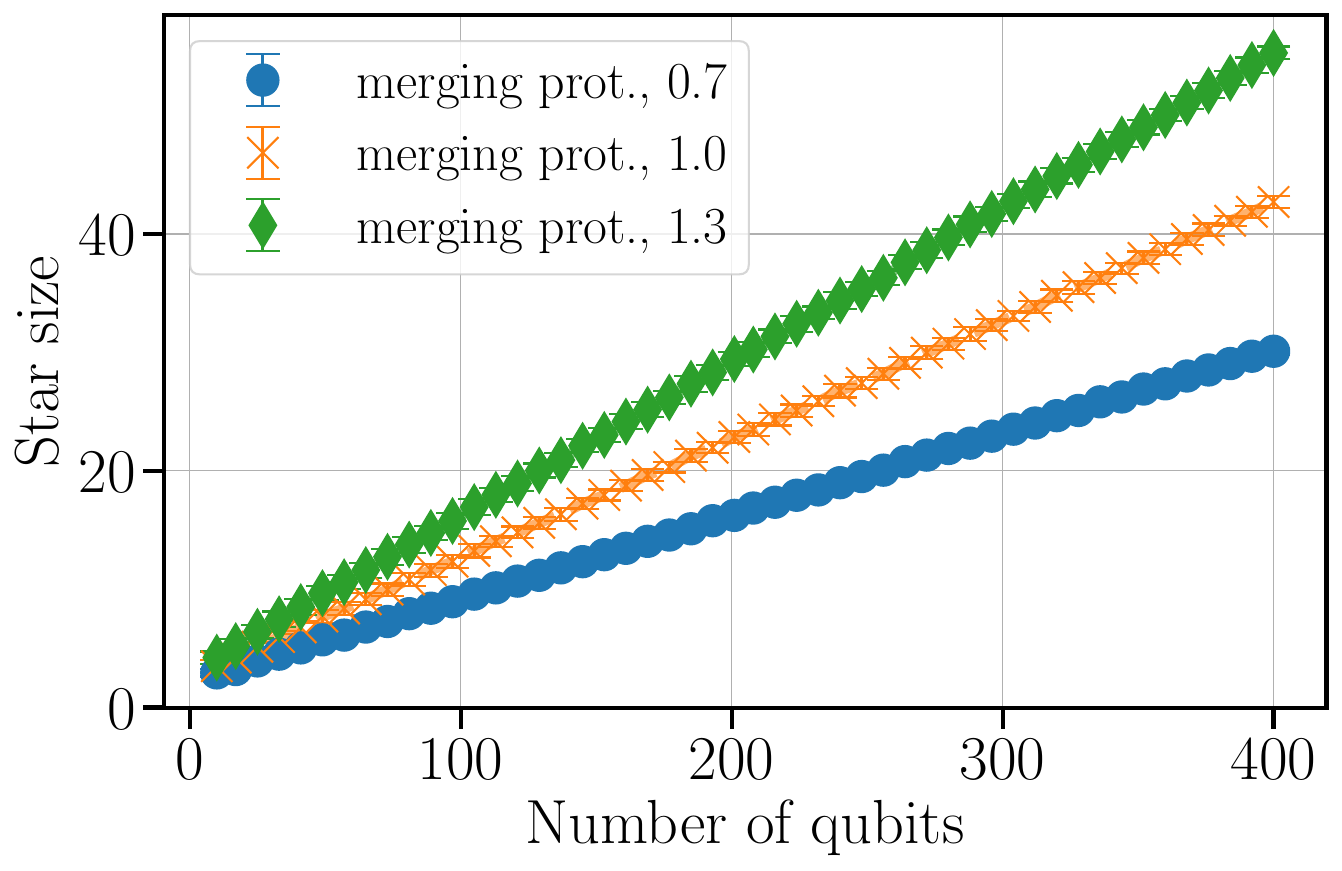}\\
    (a) & (b) & (c) \\
    \includegraphics[width=0.3\textwidth]{figures/circuit_sampling_eval_random_graph_endros_renyi_1_p0.5_circuit_depth.pdf}&
    \includegraphics[width=0.3\textwidth]{figures/circuit_sampling_eval_random_graph_endros_renyi_1_p0.5_number_measurements.pdf}&
    \includegraphics[width=0.3\textwidth]{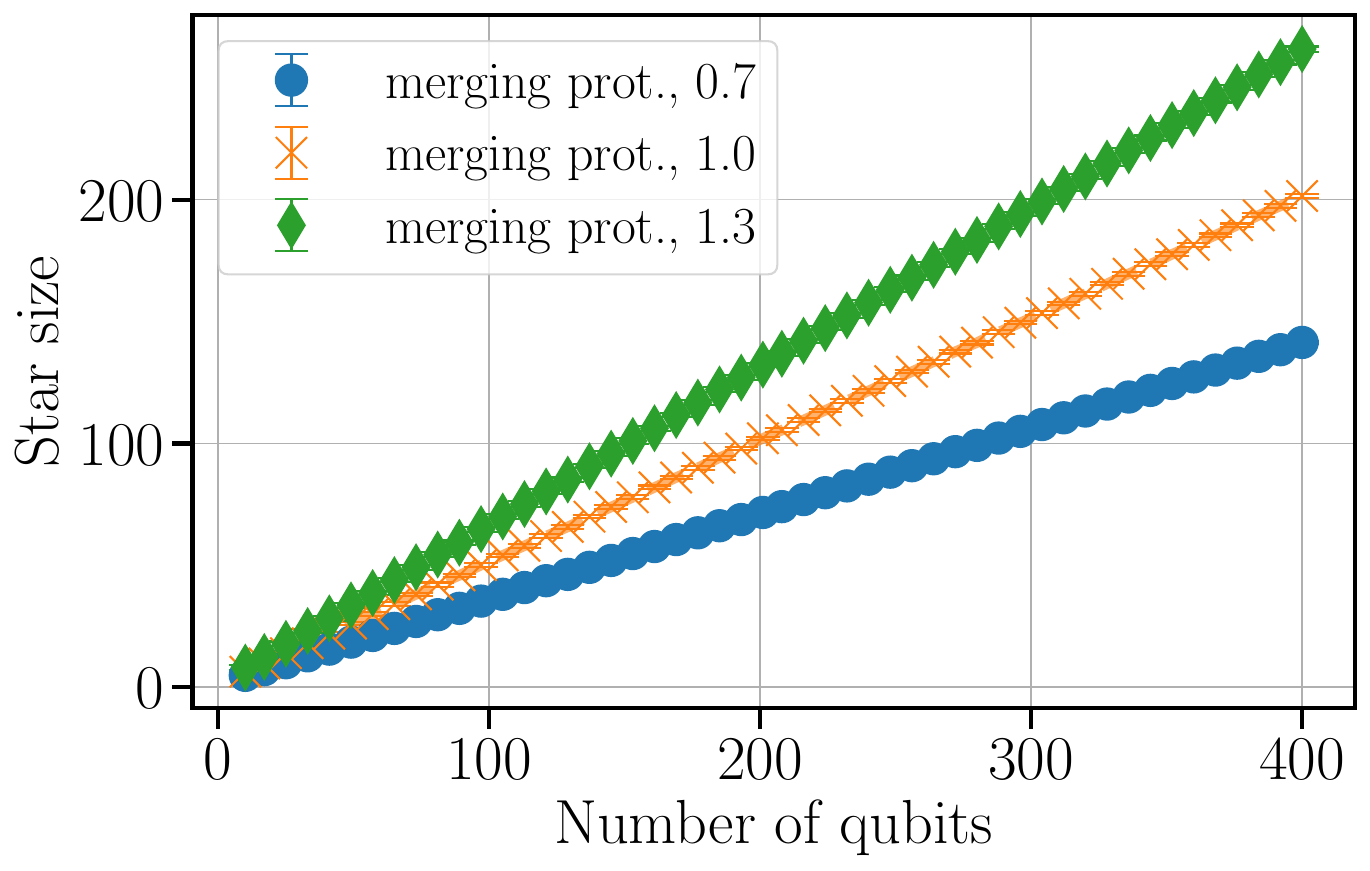}\\
    (d) & (e) & (f) \\
    \includegraphics[width=0.3\textwidth]{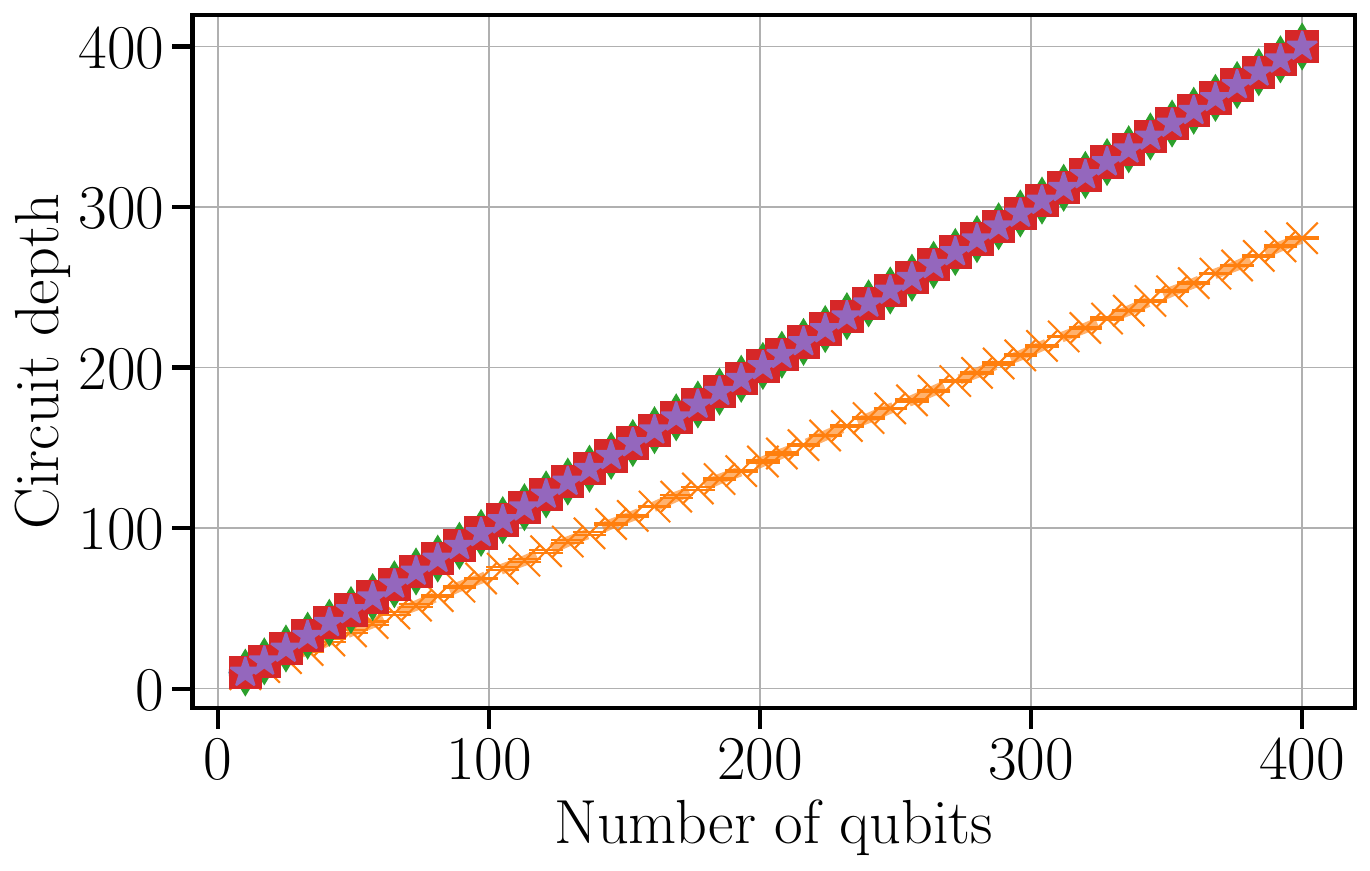}&
    \includegraphics[width=0.3\textwidth]{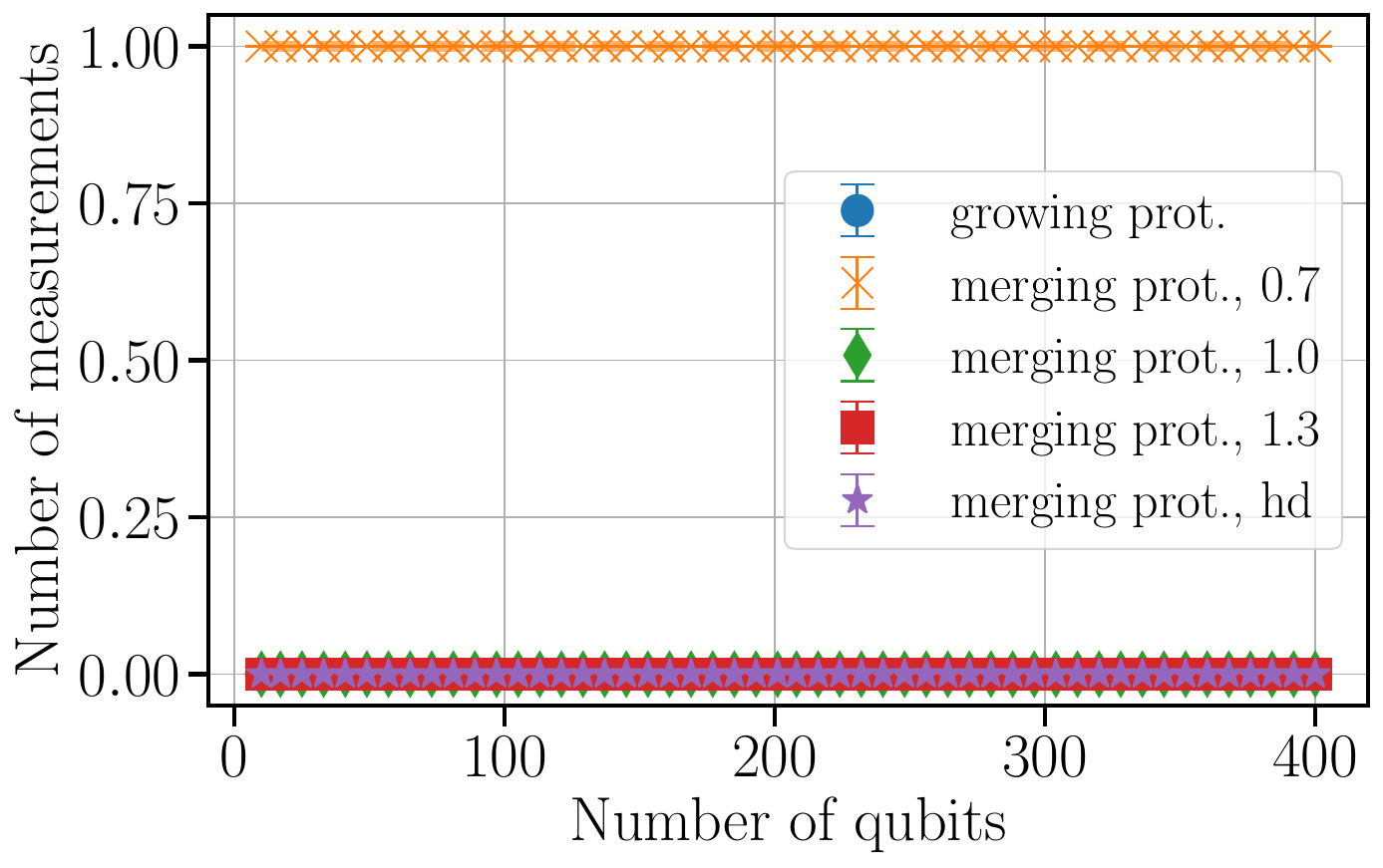}&
    \includegraphics[width=0.3\textwidth]{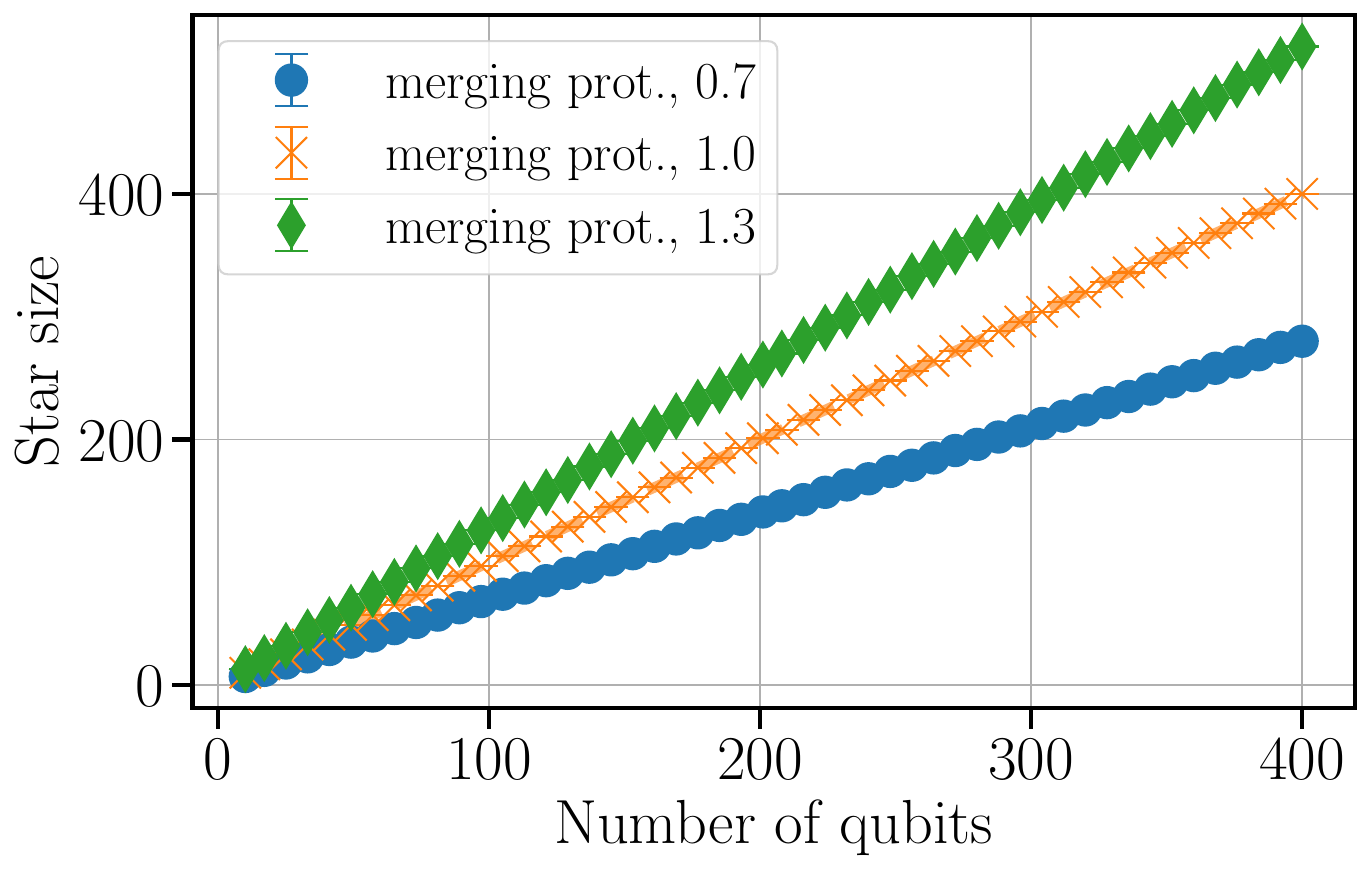}\\
    (g) & (h) & (i) \\
    \end{tabular}
    \caption{Sampling from Erd\H{o}s--R\'enyi random graphs~\cite{erdosrenyi1959,erdosrenyi1960}. Here we compare the \textit{growing} protocol with our \textit{merging} protocol using a varying target size for the star selection. The curves represent \textit{growing} protocol (blue circles) as well as the \textit{merging} protocol with a star selection based on scaling factors $0.7$ (orange crosses), $1.0$ (green diamonds), $1.3$ (red squares) and the highest degree (violet stars). As explained in Section~\ref{sec:merging}, the scaling factor relates the target star size for every star selection during \textit{merging} with the average star size in the input graph. A larger scaling factor thus corresponds to a larger star size (cf. Panels (c), (f) and (i)) The legends of panels (a) and (b), (d) and (e), (g) and (h) coincide and are thus only shown in panels (b), (e) and (h). The rows $1$, $2$, $3$ (top to bottom) show the comparison for the graph generation in the Erd\H{o}s--R\'enyi model with $p=0.1$, $p=0.5$ and $p=1.0$ respectively. The collumns $1$, $2$, $3$ (from left to right) show the averaged circuit depth, number of measurements and star size for \textit{merging} in dependence of the number of qubits in the final GHZ state, respectively. For each GHZ state size, we generated $100$ Erd\H{o}s--R\'enyi random graphs of this size. The error bars show the standard deviation obtained from averaging over those samples.}
    \label{fig:random_graph_merge_vs_grow_num_qubits}
\end{figure*}

Contrary to the IBM and rectangular layout, we cannot determine the best criteria to select stars in any random layout graph (cf. Figure~\ref{fig:random_graph_merge_vs_grow_num_qubits}). On the one hand, a small scaling factor results in a small circuit depth. Choosing a scaling factor of $1.3$, i.e., a target degree above the average degree or directly using the highest degree criteria yields a larger average circuit depth than the \textit{growing} protocol. On the other hand, larger scaling factors result in larger stars and thus in less merging processes, i.e., less measurements. Additionally, the absolute values of the circuits depth increase for increasing $p$ values of the random graph and the number of measurements decrease. This can be explained by noting that the $p$ value determines the average degree in the random graph. For increasing average degree the star sizes get larger for a fixed scaling factor. We thus conclude that the best criteria has to be chosen for each individual case separately, as a tradeoff between small circuit depth and resulting number of measurements. Note however that it is always possible in a random layout graph to artificially reduce the $p$ value by leaving out existing edges. This will decrease the circuit depth and increase the required number of measurements and can be used to find the ideal compromise.

\end{document}